\documentclass[namedreferences]{solarphysics}

\usepackage[hyperref,optionalrh,showbiblabels]{spr-sola-addons} % For Solar Physics 
\usepackage{graphicx}        % For eps figures, newer & more powerfull
\usepackage{color}           % For color text: \color command
\usepackage{breakurl}        % For breaking URLs easily trough lines
\chardef\us=`\_

\begin{document}

\begin{article}
\begin{opening}

\title{Forecasting Solar Flares by Data Assimilation in Sandpile Models.
%II. Efficient assimilation and predictive potential
\\ {\it Solar Physics}}

\author[addressref=aff1,corref,email={
christian.thibeault@umontreal.ca}]{\inits{C.}\fnm{Christian}~\lnm{Thibeault}}%\sep
\author[addressref=aff2,corref,email={antoine.strugarek@cea.fr}]{\inits{A.}\fnm{Antoine}~\lnm{Strugarek}}%\sep
\author[addressref=aff1,corref,email={paul.charbonneau@umontreal.ca}]{\inits{P.}\fnm{Paul}~\lnm{Charbonneau}}
\author[addressref=aff3,corref,email={btremblay@nso.edu}]{\inits{B.}\fnm{Benoit}~\lnm{Tremblay}}\sep
%\author[corref,email={paul.charbonneau@umontreal.ca}]{\inits{P.}\fnm{Paul}~\lnm{Charbonneau}}%\sep
%\author{\inits{}\fnm{}~\lnm{}\orcid{}}
%\author{P.~\surname{Author-a}$^{1}$\sep
%        E.~\surname{Author-b}$^{1}$\sep
%        M.~\surname{Author-c}$^{2}$      
%       }

%   \institute{$^{1}$ First affiliation
%                     email: \url{e.mail-a} email: \url{e.mail-b}\\ 
%              $^{2}$ Second affiliation
%                     email: \url{e.mail-c} \\
%             }
\address[id=aff1]{Departement de physique, Universite de Montreal, C.P. 6128 Succ. Centre-ville, Montreal H3C 3J7, Canada}
\address[id=aff2]{Departement d'Astrophysique/AIM, CEA/IRFU, CNRS/INSU, Univ. Paris-Saclay, Univ. de Paris, 91191 Gif-sur-Yvette, France}
\address[id=aff3]{Laboratory for Atmospheric \& Space Physics, 3665 Discovery Drive, Boulder, CO 80303, U.S.A.}

\runningauthor{Thibeault et al.}
% TODO CHECK THIS OUT
\runningtitle{Forecasting Solar Flares by Data Assimilation in Sandpile Models}

\begin{abstract}
The prediction of solar flares is still a significant challenge in space weather research, with no techniques currently capable of producing reliable forecasts performing significantly above climatology. In this paper, we present a flare forecasting technique using data assimilation coupled with computationally inexpensive cellular automata called sandpile models. Our data assimilation algorithm uses the simulated annealing method to find an optimal initial condition that reproduces well an energy-release time series. We present and empirically analyze the predictive capabilities of three sandpile  models, namely the Lu and Hamilton model (LH) and two deterministically-driven models (D). Despite their stochastic elements, we show that deterministically-driven models display temporal correlations between simulated events, a needed condition for data assimilation. We present our new data assimilation algorithm and demonstrate its success in assimilating synthetic observations produced by the avalanche models themselves. We then apply our method to GOES X-Ray time series for 11 active regions having generated multiple X-class flares in the course of their lifetime. We demonstrate that for such large flares, our data assimilation scheme substantially increases the success of ``All-Clear'' forecasts, as compared to model climatology.
\end{abstract}
\keywords{Flares, Forecasting, Avalanche models, Data assimilation, Self-organized criticality}
\end{opening}
%-------------------------------------------------

\section{Introduction}

The accurate prediction of large solar flares is one of the desired milestones in ongoing space weather research efforts. A number of semi-empirical, statistical, and/or data-driven techniques have been designed over the years, with various degrees of success.  Nonetheless, none of them seem so far to be consistently doing very much better than so-called climatological forecasting, which consists in simply predicting according to empirically constructed statistical distributions. The flare forecasting exercise reported upon in \cite{barnes_comparison_2016} (see also \cite{leka_comparison_2019} and \cite{leka_comparison_2019-1}) is particularly interesting in quantifying the relative merits of these various techniques.

That reliable flare forecasting should prove hard to achieve is not at all surprising. The first major challenge is related to the extremely wide range of spatial and temporal scales characterizing the flaring phenomenon. This makes brute force approaches based on magnetohydrodynamical simulations extremely challenging, if not unpractical (see \cite{cheung_comprehensive_2019} for arguably the closest yet). Empirical/statistical techniques face another related challenge: the observed size distributions of flares take the form of a steep power law (e.g. \cite{dennis_solar_1985}; \cite{aschwanden_self-organized_2011}), meaning that large flares, which are the most important to accurately forecast, are rare. Consequently, the observational statistics are dominated by small events, implying in turn that methods such as machine learning have (relatively) few examples of the largest flares available for training purposes.

The power-law shape of flare properties (peak flux, duration, total released energy) is an indication of scale invariance, which is a strong hint as to the inner workings of the flare phenomenon. Furthermore, similar power laws are constructed from stellar flare data (see \cite{aschwanden_scaling_2008} and \cite{namekata_statistical_2017}), pointing to universality in the flaring process. This observed scale invariance has led to the consideration of flares as possibly arising from an avalanche of small reconnection events cascading across coronal loops or other magnetic structures. \cite{lu_avalanches_1991} and \cite{lu_solar_1993} have designed the first lattice-based sandpile models capturing this avalanching process, with numerous variations on the theme subsequently proposed as explanatory frameworks for flares in general (for reviews, see \cite{charbonneau_avalanche_2001} and \cite{aschwanden_25_2016}). A key aspect of the Lu \& Hamilton proposal is that their lattice, an idealized representation of a coronal structure loading magnetic energy, autonomously reaches a critical state in response to slow external forcing, from which results scale-invariant impulsive energy release. This represents an instance of self-organized-criticality (hereafter SOC; see \cite{bak_self-organized_1987} or \cite{jensen_self-organized_1998}), now understood to be a robust generator of scale-invariant behavior, including power-law size distribution for energy release events in general. Indeed, the SOC framework has been applied to phenomena as diverse as solar flares, earthquakes, lightning, and geomagnetic substorms, to name but a few \citep{aschwanden_self-organized_2011,2016SSRv..198....3W}.

In the context of solar flare, the SOC hypothesis is buttressed by a physical scenario due to Parker \citep{parker_nanoflares_1988}, according to which photospheric fluid motions twist and braid the footpoints of magnetic fieldlines within coronal loops, leading to the inexorable buildup of magnetic tangential discontinuities (or current sheets) becoming unstable and releasing thermal energy in the form of what Parker dubbed nanoflares. Although originally designed as a model for coronal heating, Parker's scenario contains all required elements for scale-invariant release of magnetic stress by cascades of localized reconnection events within a coronal loop. This is the physical picture captured by the sandpile models of the general type introduced by Lu \& Hamilton.

All avalanche models proposed in the flare context involve stochastic elements, either in lattice loading, avalanche triggering, and/or internal redistribution in the course of avalanches. This may lead one to expect that such models, even if they properly capture statistical flaring behavior, should be useless for prediction of individual events. For any of the aforecited sandpile models, the triggering and unfolding of the numerous small avalanches, equivalent to the more frequent, smaller flares, are indeed strongly affected by the stochastic elements embedded in the model, and so are truly unpredictable.

However, and perhaps counter-intuitively, this may not be the case for the larger avalanches. In sandpile models, large avalanches release large-scale stress patterns having built up in the lattice in response to slow external forcing, but also via the unfolding of earlier avalanches, especially the larger ones. Once they reach the self-organized critical state, sandpile models exhibit long range spatiotemporal correlations \citep{jensen_self-organized_1998}, so that even in the presence of truly stochastic elements, the system does not behave completely stochastically. What this implies is that information useful for predicting future avalanching behavior is contained, in principle, in past avalanching behavior. How to extract that information then becomes the key challenge.

Data assimilation is one attractive possibility. In short, the idea is to use past observations to adjust the state of an underlying physical model to best reproduce these observations, yielding an ``optimal'' initial condition for model-based forecasting. Such procedures have been extensively used for (earthly) weather forecasts. \cite{belanger_predicting_2007} and \cite{strugarek_predictive_2014} have presented proof-of-concept and exploratory results on the use of various data assimilation methods in the context of sandpile models for flares. These latter authors, in particular, have shown an example of successful assimilation of both synthetic data as well as GOES X-Ray flux time series. This prior work is the motivation and starting point of the work reported upon here. 

In this paper, we first introduce in section \ref{sec:AvalModels} the three sandpile models used in our analysis, namely the Lu and Hamilton model (LH) and two deterministically driven models (D), taken from
\cite{strugarek_predictive_2014}. In section \ref{sec:PrefCapAvalModel}, we present empirical analyses and comparison of the predictive potential of these three models. In section \ref{sec:DAarticle}, we present our data assimilation protocol and our results in assimilating synthetic observations produced by the sandpile models.
In section \ref{sec:GoesDA} we apply our assimilation scheme to a set of GOES X-Ray flux time series, and demonstrate that data assimilation yields a substantial improvement in forecasting skill for large avalanches, over and above model climatology. We conclude in \ref{sec:conclusion} by summarizing our method and results, and outlining the path to future improvements.

\section{Sandpile models for solar flares}
\label{sec:AvalModels}
Sandpile (or avalanche) models are cellular automata that can be used to represent the flaring solar corona in a highly simplified view. Despite their simplicity, they capture at some level the threshold dynamics and disparity of spatial and temporal scales characterizing the flaring phenomena, and reproduce reasonably well the power-law size distributions of solar flares. They also draw physical support from the nanoflare picture proposed by \cite{parker_nanoflares_1988}. In comparison to MHD models of coronal magnetic fields, sandpile models are computationally inexpensive and thus, in principle, can be more easily coupled with data assimilation methods. For detailed reviews on sandpile models for solar flares, we refer the reader to \cite{charbonneau_avalanche_2001} and \cite{aschwanden_25_2016}. 

\subsection{The Lu and Hamilton Model}

The first sandpile model of solar flares was published by Lu and Hamilton in 1991 \citep{lu_avalanches_1991}. As such, it is often considered as the canonical avalanche models for solar flares, which we denote as ``LH" hereafter. Here, we use a 2D Cartesian lattice version of this model, with $48 \times 48$ nodes each assigned a real number denoted $A^n_{i,j}$, where $n$ is a discrete time index and $(i,j)$ are its lattice coordinates. The driving of the lattice takes place via addition of a small increment value $\delta A$ on a single randomly selected node, at every non-avalanching temporal iteration. The value of $\delta A$ is randomly selected through a uniform distribution between $-0.2$ and $0.8$. The boundary of the lattice is kept at $A=0$. 

The stability of a node is determined by its local curvature, approximated as:

\begin{equation}
\Delta A^n_{i,j} \equiv A^n_{i,j} - \frac{1}{4}\sum_{k=1}^4 A^n_k ~,
    \label{StabilityLH}
\end{equation}
where $k$ denotes each nearest neighbor of the node $(i,j)$. Once $|\Delta A^n_{i,j}|$ exceeds a predetermined threshold $Z_c$, driving is suspended and redistribution occurs to restore stability. This is achieved by transferring an amount of nodal value from the unstable node $(i,j)$ to each neighbor according to the following rules:

\begin{equation}
A^{n+1}_{i,j} = A^n_{i,j} - \frac{4}{5} Z ~,
    \label{RedistributionLH}
\end{equation}
\begin{equation}
A^{n+1}_{i\pm1,j\pm1} = A^n_{i\pm1,j\pm1} + \frac{1}{5} Z ~,
    \label{RedistributionLH_2}
\end{equation}
where $Z \equiv Z_c \Delta A^n_{i,j}/|\Delta A^n_{i,j}|$. This process is deterministic and conservative, meaning that the total amount of nodal value is conserved during the redistribution. This redistribution can cause one or more neighbouring nodes to exceed the stability threshold, in which case redistribution is applied anew to these unstable nodes, and so on until stability is everywhere restored. Only then does driving resumes (``stop-and-go'' sandpile). These chain reactions of redistribution events are what we call \textit{avalanches}, and represent the model equivalent of a solar flare. 

Since the nodal value $A^n_{i,j}$ is generally associated with a measure of the magnetic twist in a coronal loop, the usual measure of the magnetic energy of the lattice is defined by: 

\begin{equation}
E^n = \sum_{i,j} A_{i,j}^2 ~,
\label{LatticeEnergy}
\end{equation}
where we sum on all nodes of the lattice.
Each redistribution will therefore cause a decrease in lattice energy by the amount:

\begin{equation}
\Delta e^n_{i,j} = \frac{4}{5 e_0} \left(  2 \frac{|\Delta A^n_{i,j}|}{Z_c} -1  \right) Z_c^2 ~,
\label{Energyrelease}
\end{equation}
where $e_0$ is the energy liberated by a single node exceeding the stability threshold
by an infinitesimal amount, expressed as:

\begin{equation}
\label{Eq-e0}
e_0 = \frac{4Z_c^2}{5} ~.
\end{equation}

In all that follows, we use a threshold value $Z_c=2$, and the above quantity as the unit to express energy released
by avalanches of all sizes.

The total energy released by the lattice per iteration ($\Delta E^n_r$) is equal to the total energy loss of all redistributing nodes. Note that other physical interpretations of the lattice nodal variable can lead to other measures of energy release in avalanche models (for a more in-depth discussion, see chapter 12 in \citealt{aschwanden_self-organized_2013} and \citealt{farhang_energy_2019}).

Sandpile models describe the evolution of a system on two characteristic timescales, the forcing timescale corresponding to the typical evolution time of an active region, of the order of hours to days, and the eruption timescale corresponding to the impulsive flare of the order of seconds. In what follows, the total energy released by an avalanche in a sandpile model will be directly compared to the peak energy flux of a solar flare in the X-Ray domain, as the subsequent  cooling of the flaring region is not accounted for in the sandpile model.

\subsection{The Deterministically Driven Sandpile Model}

A deterministically driven model in the context of solar flare simulations was first proposed by \cite{strugarek_deterministically_2014}, notably inspired by deterministic driving schemes used for seismic faults \citep{olami_self-organized_1992} and geomagnetic substorms \citep{liu_energy_2006,vallieresnollet_dual_2010}. The lattice is driven globally, through the following deterministic forcing rule:

\begin{equation}
A_{i,j}^{n+1} = A_{i,j}^n \times (1 + \epsilon ) , \; \; \epsilon \ll 1, \; \; \forall (i,j) ~,
\label{DrivingDet}
\end{equation}
where $\epsilon \ll 1$ is the driving rate. We use a value $\epsilon=10^{-5}$ and denote these models as ``D" in what follows.

 The stability criterion remains the same as in the LH model (equation (\ref{StabilityLH})): once the curvature of a node exceeds a threshold $Z_c$, redistribution occurs evenly among all neighbors as in the LH models, but with random loss of nodal value induced per redistribution:

\begin{equation}
A^{n+1}_{i,j} = A^n_{i,j} - \frac{4}{5} Z ~,
    \label{RedistributionDet}
\end{equation}
\begin{equation}
A^{n+1}_{i\pm1,j\pm1} = A^n_{i\pm1,j\pm1} + \frac{r_0}{5} Z ~,
    \label{RedistributionDet_2}
\end{equation}
where $r_0$ is extracted from a uniform distribution spanning $[D_{nc},1]$. This redistribution rule is non-conservative, as each redistribution now involves a random loss of nodal value from the lattice. As a consequence, the stochastic element enters the unfolding of avalanches, rather than in the driving mechanism as in the LH model.

As argued in \citet{strugarek_deterministically_2014}, the deterministic driving introduced above can be (loosely) interpreted as a global twisting of a coronal loop by photospheric flows coherent on scales commensurate with the loop's diameter. In what follows we consider two versions of deterministically-driven models encapsulating different levels of stochasticity (corresponding to model D2 and D3 in \citealt{strugarek_predictive_2014}). The first, hereafter dubbed Model D09, uses a conservation parameter $D_{nc} = 0.9$ and is considered a mildly non-conservative model. The second, Model D01, is a more strongly non-conservative model, with $D_{nc} =0.1$, and thus introduces a larger degree of stochasticity in the unfolding of its avalanches.

%\textcolor{red}{
%{\bf  Content of following par should go in conclusion; here it is a distraction I think} 
%Another class of deterministically-driven models for solar flares has been recently proposed by \cite{farhang_principle_2018}, where the liberated energy of each redistribution is maximized to mimic energy minimization during magnetic reconnection.Here, we only explore the initial model of \cite{strugarek_predictive_2014} and defer the detailed analysis of this new class of deterministic models for future works.}   \AS{ok for me}

\subsection{Avalanching properties of LH and D models}

\begin{figure}
    \centering
    \includegraphics[width=1\linewidth]{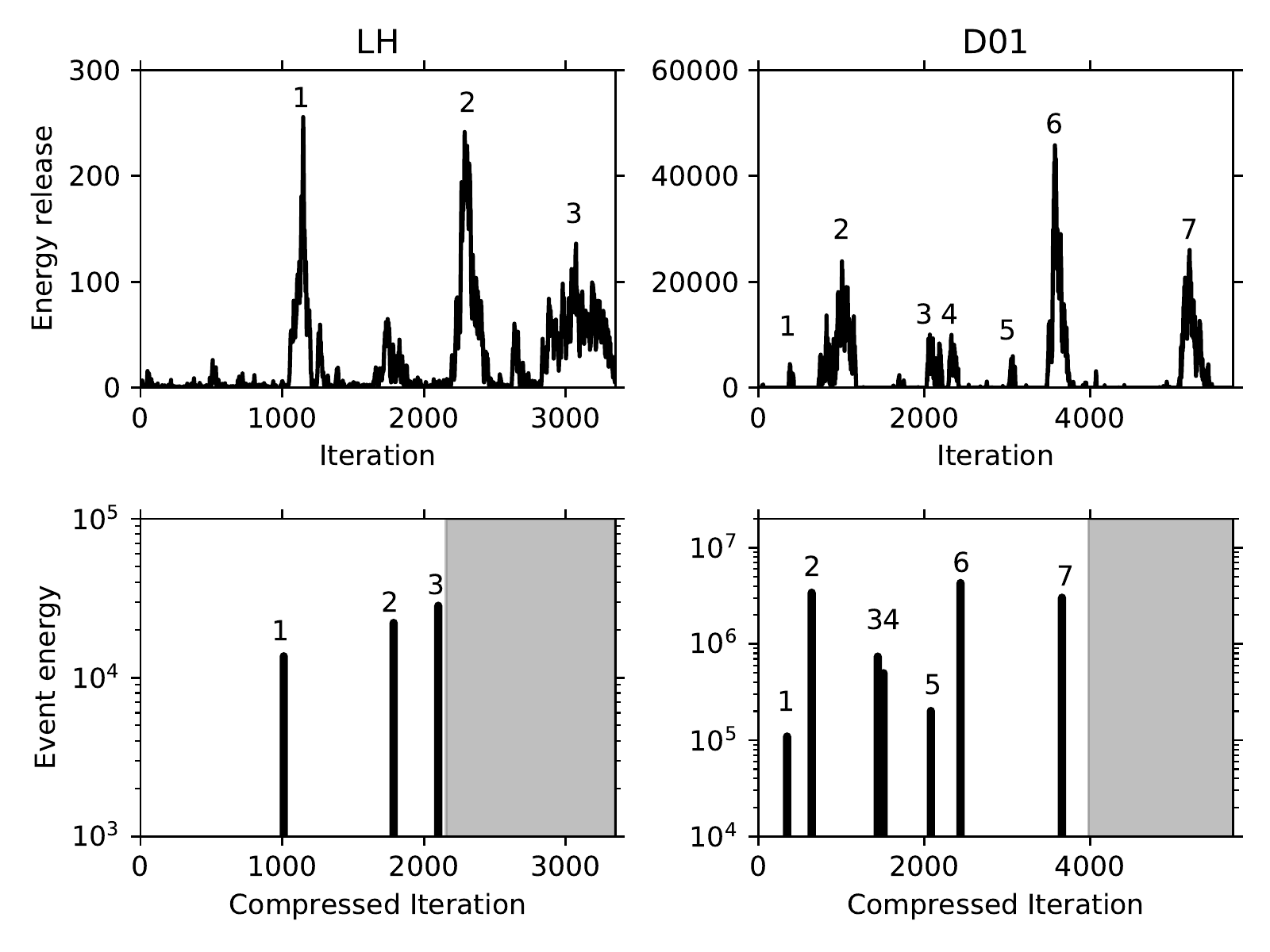}
     \centering
    \caption[Top: Energy release time series for model LH (left) and D01 (right). Bottom: Compression of the same energy release time series, where each line represents the total energy released during the avalanche.]{Top: Energy release time series for model LH (left) and D01 (right). Bottom: Compression of the same energy release time series, where each line represents the total energy released during a large avalanche (over $10^5$ for model D01 and $10^4$ for the LH model). 
    Here as on all subsequent plots showing energy release time series, energy is measured in units of $e_0$ (viz.~eq.~(\ref{Eq-e0})). Individual avalanches are labelled in each panel (1 to 3 for left panels, 1 to 7 for right panels). The grey area is added to represent the iterations lost during the compression algorithm. Compressed avalanches below threshold are not numbered.}
        \label{EnergyReleaseCompression}
\end{figure}

We illustrate typical energy release time series for models LH and D01 in the top panels of Figure \ref{EnergyReleaseCompression}. Model D01 clearly tends to produce much larger avalanches than model LH.
%a direct consequence of the deterministic driving mechanism that adds stress simultaneously to all nodes. 
To explicitly account for the large separation of timescales between the release of energy in solar flares and their waiting times, we follow \cite{lu_solar_1993} and compress avalanches into discrete events (as shown in the bottom panels of Figure \ref{EnergyReleaseCompression}), each representing the sum of all energy released during an avalanche. 
%\cite{strugarek_predictive_2014} have shown that D models hold better predictive capabilities for large avalanches. T
Since we wish to focus on large avalanches, we only retain avalanches preceding a preset threshold, which we (somewhat arbitrarily) set at $10^5$ for models D01 and D09, and $10^4$ for the LH model.

\begin{figure}
    \centering
    \includegraphics[width=0.75\linewidth]{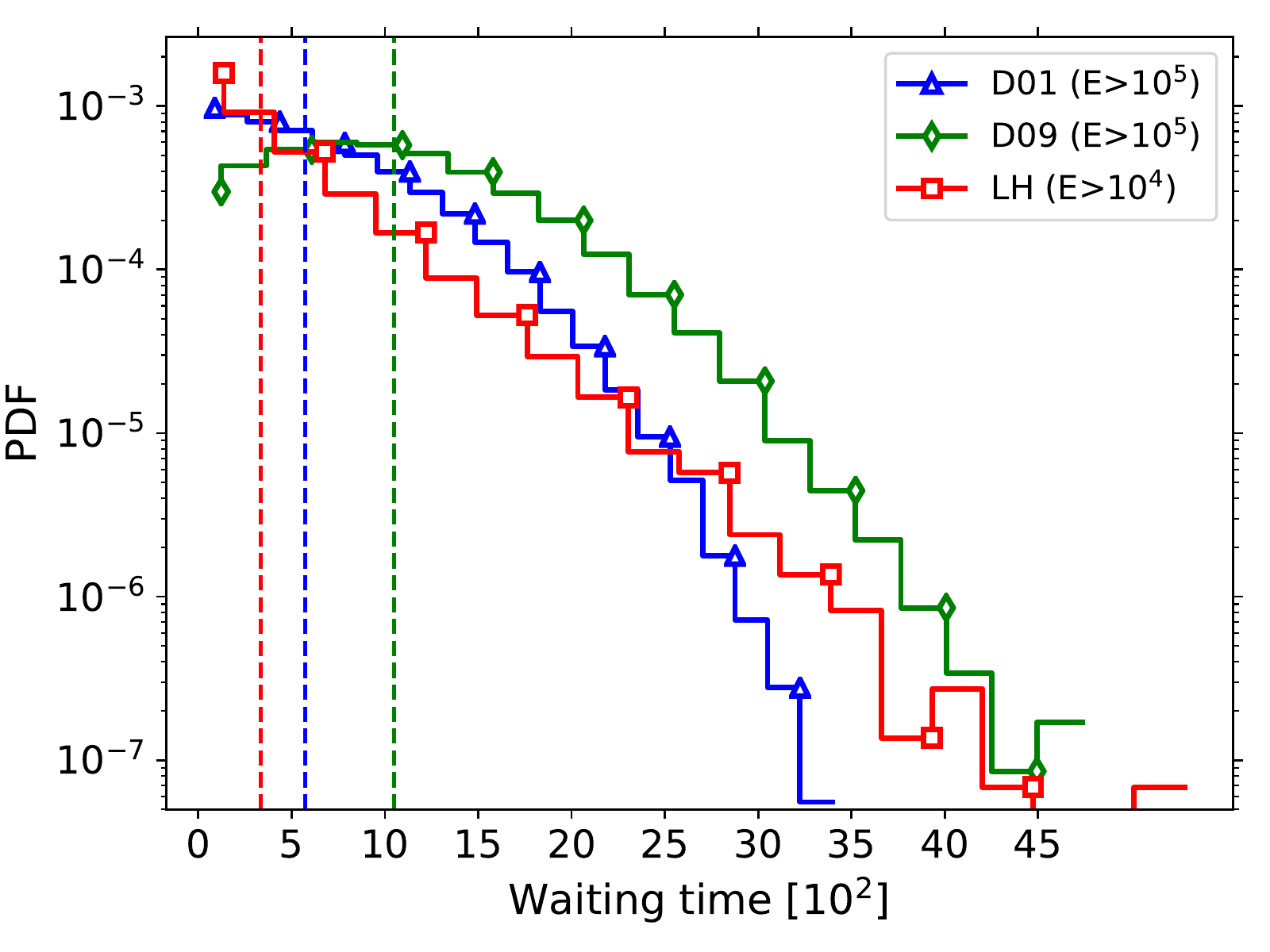}
     \centering
    \caption[Description.]{Waiting time distribution between avalanching events releasing energy over 10$^4$ (in $e_0$ units) for Model LH and over 10$^5$ for Model D01 and D09, with median waiting times shown by dashed vertical lines.}
        \label{WT_stats}
\end{figure}

%Figure \ref{WT_stats} shows histograms of waiting time between events of all sizes (blue, top) and between larger events (red, bottom) for models LH and D01. In both models, a clear exponential distribution is seen for waiting times of all events, reminiscent of a random Poisson process \citep{wheatland_origin_2000}. This result is expected for model LH, which triggers avalanches using a random driving process, although more surprising for model D01 with a deterministic triggering of avalanches. For larger events, model D01 departs from this exponential distribution, reminiscent of a load-unload cycle \citep{rosner_cosmic_1978}. This is not the case with model LH, which maintains its exponential distribution even for higher events.   
Figure \ref{WT_stats} shows histograms of waiting time between events above a given threshold for models LH, D09 and D01. In model LH, a clear exponential distribution is seen, indicative of a stationary Poisson process \citep{wheatland_origin_2000}. This result is expected for model LH, which triggers avalanches using a random driver. In the D models, the distribution could be approximated to an exponential law for intermediate-size events. For larger events, model D01 departs from this exponential distribution. %, reminiscent of a load-unload cycle \citep{rosner_cosmic_1978}. This is not the case with model LH, which maintains its exponential distribution even for higher events.   

%\textcolor{red}{
%{\bf le paragraphe qui suit pourrait aussi aller dans la conclusion, ici c'est encore une distraction je trouve; vos opinions ?}
%We recognize that our exponential distributions do not match the power-law waiting times distributions observed from solar flares \citep{wheatland_origin_2000}. Other published sandpile models have addressed this problem by applying a non-stationary driving mechanism \citep{norman_waiting-time_2001}, or more recently, an energy minimization method during the redistribution of unstable nodes  \citep{farhang_principle_2018}. Since this work concentrates on the predictive capabilities of avalanche models and our data assimilation algorithm, we chose to address this problem in future work.}

\section{Predictive capabilities of avalanche models}
\label{sec:PrefCapAvalModel}
\subsection{Robustness of large avalanches}
\label{RobustLarge}

\begin{figure}
    \centering
    \includegraphics[width=0.75\linewidth]{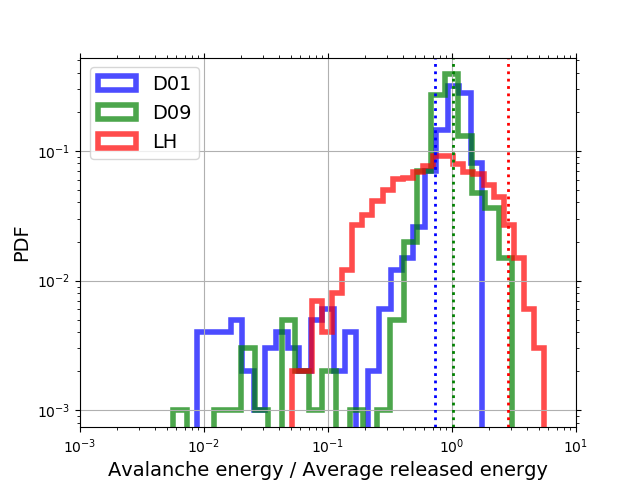}
    \includegraphics[width=0.75\linewidth]{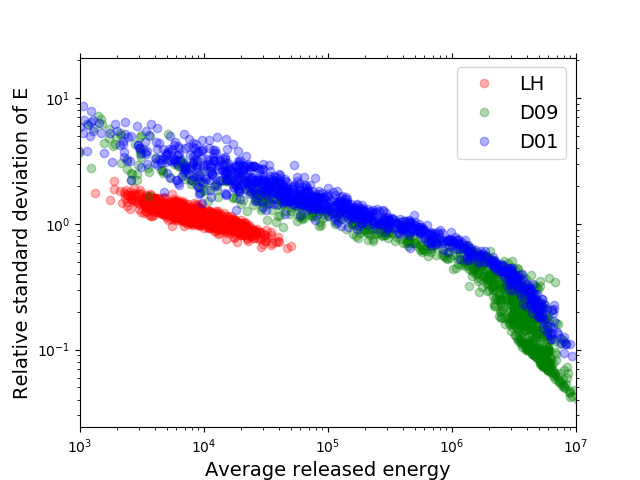}
     \centering
    \caption[Top: Probability density function (PDF) of energy released from 1000 avalanches produced from the same initial condition, but with varying random-number sequences. Bottom: Relative standard deviations of 1000 distributions of energy released (E) such as the one presented in the top panel of this figure, with respect to the average energy released of each distribution.]{Top: Probability density function (PDF) of energy released from 1000 avalanches produced from the same initial condition, but with varying random-number sequences. The initial condition is chosen to precede a large avalanche. 
    Bottom: Relative standard deviations of 1000 distributions of energy released (E) such as the one presented in the top panel of this figure, with respect to the average energy released of each distribution.}
        \label{DistE}
\end{figure}

\cite{strugarek_predictive_2014} showed with a limited exploration that deterministically driven models have a strong potential to robustly predict large events compared to LH models. We further expand their analysis to systematically characterize this ability. From one fixed initial condition preceding a large avalanche, we used 1000 different random-number sequences to drive the LH model until an avalanche is triggered. Likewise, we use 1000 distinct random number sequences to control avalanching behaviour in the deterministically-driven D01 and D09 models. The distribution of energy released in the first avalanche in all the 1000 runs is presented in the top panel of Figure \ref{DistE}. 
In this example, models D01 and D09 both have a very high tendency to produce large avalanches, regardless of their random-number sequences. This suggests that a pre-existing stress pattern in the lattice is an important factor in producing a large avalanche. In the case of the LH model, the distribution is much wider, indicating a larger sensitivity to the random-number sequence in triggering a large avalanche. 

We then investigated how these distributions of events change with respect to different initial conditions preceding avalanches of different sizes. To do so, we produced 1000 distributions of energy released from initial conditions preceding large avalanches, such as the ones presented in top panel of Figure \ref{DistE}. With each of these distributions, we extract the average and relative standard deviation of the energy release. The bottom panel of Figure \ref{DistE} shows the relationship between these two extracted parameters. In general, the distribution of events tends to be more concentrated as the average energy released increases for all models. The cutoff of the LH model to events higher than $10^5$ is a characteristic of the model itself, which tends to produce smaller events. For the D models, the standard deviation of the distributions drops for events higher than $10^6$, especially in the case of model D09. This confirms previous findings that D models tends to favor large avalanches in its predictive capabilities.

\subsection{Short-term correlations between large events}

The results shown in the previous section already suggest that the timing of a large avalanche is dictated by the stress patterns established across the lattice by previous avalanches. 
To assess whether data assimilation can be successfully coupled to sandpile models, we verified that these models also contain short-term correlations between large events. We also characterized the memory timescale associated with stress patterns.
%We have intuited in section \ref{RobustLarge} that the timing of a large avalanche is dictated by the stress patterns established across the lattice by previous avalanches. 
%Although, we have not characterized the memory timescale associated with these stress patterns.
Towards this end, we first construct a composite time series by superposing the compressed energy release time series
of 10 000 model runs, each using the same  initial condition but with a distinct stream of random-number for driving (LH) or redistribution (D01, D09). Any large peak in such a composite time series is indicative of preferred time for avalanching, i.e., memory.

\begin{figure}
    \centering
    \includegraphics[width=1\linewidth]{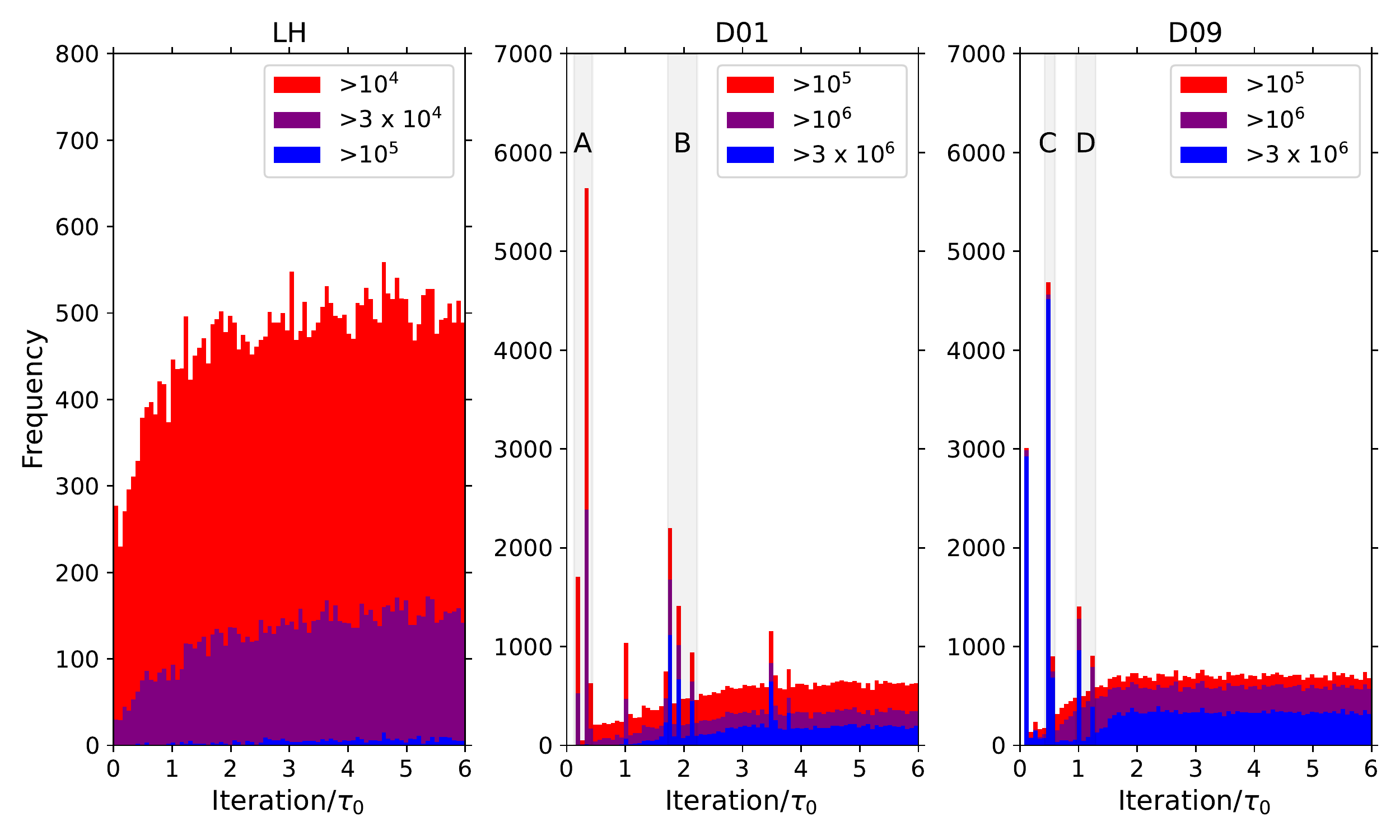}
     \centering
    \caption[Occurence time for large avalanching events (energy over $10^4$ for the LH model and over $10^5$, for both D models) from 10 000 models runs, each with identical initial conditions but with varying random-number sequences.]{Occurence time for large avalanching events (over $10^4$ for the LH model and over $10^5$ expressed as before in units of $e_0$, for both D models) from 10 000 models runs, each with identical initial conditions but with varying random-number sequences. The grey highlighted areas, labeled as A, B, C and D are time windows used to analyse correlations between the timing of large events.}
        \label{EventTimingDistribution}
\end{figure}

Such composite time series are shown in Figure \ref{EventTimingDistribution}, for models
LH (left), D01 (middle) and D09 (right). In all cases,
the time axis is normalized by the median waiting time ($\tau_0$) between events larger than $10^5$ for D models and $10^4$ for the LH model (these characteristic times are also indicated in Figure \ref{WT_stats}). In the case of the LH model (leftmost panel of Figure \ref{EventTimingDistribution}), we see no preferential timing of large events. These results are consistent with previous findings regarding the low predictive capabilities of the LH model  \citep{strugarek_predictive_2014}. For the deterministically driven model D01 (middle panel), the results show clear peaks in the timing of large events. These peaks  cluster within time windows smaller than $\tau_0 / 2$ and persist until up to $4 \tau_0$. It suggests that the lattice contains stress patterns that tend to produce large avalanches at specific times. 
%\textcolor{red}{
%{\bf la phrase suivante m'embête un peu; la véritable première avalanche pourrait bien être sous notre seuil... faut y repenser un peu}
%The lack of avalanches prior to the very first peak is not too surprising, since the timing of the first avalanche, regardless of its size, is completely determined by the initial conditions.}
The lack of avalanches prior to the very first peak shows here that avalanches of smaller energy than the threshold and occurring before the first peak do not affect significantly its timing.

Further analysis shows that $58 \%$ of D01 simulations that produced large events ($>10^5$) in the B window (from $1.75 \tau_0$ to $2.15 \tau_0$) also produced large events in the A window, the initial phases of the simulation. Model D01 therefore shows short-term correlations between events, meaning that large events produced in one time window will coherently affect the timing of future events over a time interval spanning up to 4 median wait time. In the case of model D09, the peaks are composed mainly of very large events ($> 3 \times 10^6$, in blue). They persist on a shorter time window of around $1.2 \tau_0$. This model also shows a lower correlation between groups of large events. For example, only $33 \%$ of runs that produced large events in the D time window also produced large events in the C window. This shows a lower and less consistent correlation between the timing of large events, in comparison with model D01. 

In summary, we have shown significant differences between the ``memory'' in our three avalanche models, even though all generate power-law PDFs with similar logarithmic slopes for their size measures, as well as roughly similar waiting time distributions. We have also shown that significant short-term correlations exist between large avalanches for Model D01. Conversely, the D09 and LH models show little to no short-term correlations between large-events. The timing distributions shown in Figure \ref{EventTimingDistribution} are representative of tens of such analyses from different initial conditions, and thus represent robust features of each model. Since the LH model does not show promising predictive capabilities, we chose to omit this model in its coupling with data assimilation in what follows\footnote{\citet{2020SoPh..295..155M} show that lattice mass (or energy) is in fact a good predictor of large avalanche in the LH model. However, this quantity is not directly accessible observationally, and is thus of limited value for operational forecasting purposes.}.

\section{Data assimilation}
\label{sec:DAarticle}
Data assimilation consists of using a set of external data to alter the internal state of a model until it reproduces the external data satisfactorily. It can be used to derive the most realistic parameters of a reduced model, to test if a given model can accommodate a given set of data, or to develop a physical model compatible with a given series of observations that can then be used to carry out predictions (\textit{e.g.} \citealt{hung_variational_2017}). We are interested in this paper in the latter application of data assimilation for large and rare solar flares.

The use of data assimilation to fit a model to observational data has been extensively used in many fields, notably in meteorological forecasting \citep{kalnay_atmospheric_2002}. The general idea is to fit a simulated time series to a time window of observational data by adjusting initial conditions of the model. To achieve this, gradient descent methods, such as 4D-var, are typically used to minimize a cost function $\mathcal{J}$ that measures the distance between the observed and modelled time series. In the case of avalanche models for solar flares, data assimilation is particularly challenging due to: (1) the strongly nonlinear relationship (due to threshold dynamics) between the model's internal state and output, (2) the stochastic and discrete nature of the model output, and (3) the degenerate nature of the model-to-output relationship, i.e., many distinct lattice configurations can yield similar avalanching behavior. We present in the following section an overview of the data assimilation procedure we have designed to address these challenges.

\subsection{Data assimilation Procedure}
\label{sec:DAProc}
\subsubsection{Data to assimilate}

The data to be assimilated consists of a series of discrete events characterized by an amplitude and an occurrence time (see e.g. bottom panels in Figure \ref{EnergyReleaseCompression}). The first type of data we will assimilate is \textit{synthetic data} (see \S\ref{sec:assimilSynthetic}), produced independently by the sandpile model. Synthetic data can be used to validate the data assimilation procedure with data that we know can be produced by the model itself. More specifically, we use synthetic data to determine optimal numerical values for the various parameters defining our cost function and controlling the minimization algorithm. We defer assimilation of GOES X-Ray flux data to \S\ref{sec:GoesDA}.

\subsubsection{Comparing two sets of discrete events: design of an efficient cost function}
\label{S-cost}

The fundamental quantity we want to assimilate over a preset temporal window (hereafter the {\it assimilation window}) is a series of discrete events, which call here \textit{reference data}. The output of the avalanche model is also a series of discrete events, which we refer to as \textit{model data}. Simple least-squares minimization is impractical due to the discrete nature of these data. We thus developed a dedicated cost function $\mathcal{J}$ designed to compare the distance between two series of discrete events \citep{belanger_predicting_2007}. 

\begin{figure} 
\centerline{\includegraphics[width=0.75\linewidth]{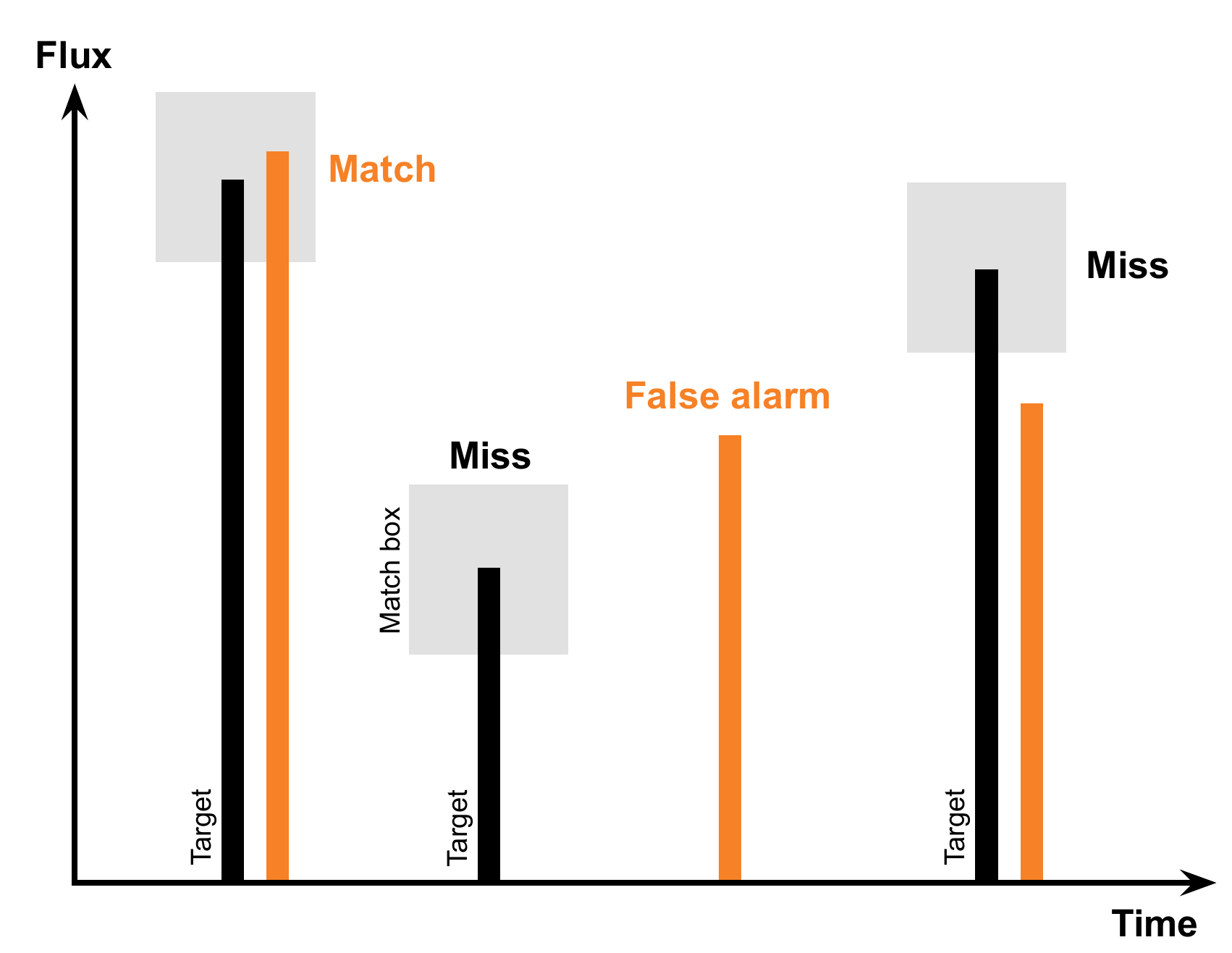}}
\caption[Illustration of our definition of a match, miss and false alarm used in our cost function between the reference (shown in black) and model data (shown in orange).]{Illustration of our definition of a match, miss and false alarm used in our cost function between the reference (shown in black) and model data (shown in orange). The gray squares represent areas in which the model data is considered a match.}
\label{fig:CostExample}
\end{figure}

Our primary goal is to accurately reproduce the largest events in a given time series. We also do not want to miss any large events, as these set the long-range spatial correlations across the lattice that confer predictability to the model, despite its stochastic elements. Finally, false alarms also need to be avoided during the assimilation process, as they will likely decrease the predictive potential of the assimilated model. For each event in the reference data and in the model data, we check whether it is a \textit{match}, a \textit{miss} or a \textit{false alarm}. Two events are considered to match when the model event occurs within a small time-window (here 400 model iterations) around the reference event, and with an energy in $[ E_0/2, 2 E_0]$ (where $E_0$ is the energy of the reference event). We illustrate the distinction between matched, missed and false alarm events in Figure \ref{fig:CostExample}. Note that our match criterion is quite demanding, any event outside the gray boxes is considered a miss, including the rightmost pair of observation and assimilated events despite their similar timing.

The cost function $\mathcal{J}$ is then defined as
\begin{equation}
  \label{eq:CostFun}
  \mathcal{J} = 1 - \left( \alpha \sum_{\mbox{match}} \frac{E_{0}}{E_{\rm tot}} -
    \beta \sum_{\mbox{miss}} \frac{E_0}{E_{\rm tot}} - \gamma
    \sum_{\mbox{False Alarm}} \frac{E_i}{E_{\rm tot}}\right)  ~,
\end{equation}
where $E_{\rm tot} = \sum E_0$ is the total energy released in the reference data over the assimilation window, $E_i$ is the energy of events in the model data, and $\alpha$, $\beta$ and $\gamma$ are adjustable weights that we set here to $1$, $0.5$ and $0.25$. This cost function thus gives more weight to large events that we want to assimilate and predict most reliably. It also gives more weight to matching events ($\alpha = 1$) than missed events ($\beta = 0.5$), and finally less weight to false alarms ($\gamma = 0.25$). If each of the reference data is matched with a model data within the match box, $\mathcal{J}=0$. Conversely, if none of the reference events are matched by the model events, $\mathcal{J} \sim 1.75$ (no match, all data events are considered as false alarms and all reference events as misses).

The cost function (\ref{eq:CostFun}) is only one example of how to design a cost function to compare two discrete series of events. We tested various flavors of $\mathcal{J}$, \textit{e.g.} by varying the coefficients $\alpha$, $\beta$ and $\gamma$ and by using the logarithm of the fractional energy of events $E_0/E_{\rm tot}$ and did not find any significant change in the results presented here.

\subsubsection{Minimization of the cost function}
\label{sec:minim-cost-funct}

In order to minimize the cost function (\ref{eq:CostFun}), several optimization techniques can be used. Before selecting a minimization algorithm, one must beforehand define which part of the model is allowed to evolve during the process. Here, we do not want to alter the avalanche model's internal parameters (threshold $Z_c$ driving rate $\epsilon$, etc.) in order to maintain solar-like statistics. The only option left is then the initial condition of the sandpile, which is typically of size $N=N_x \times N_y \sim 10^3$. Such a large set of control parameters slows down the assimilation process. In this work, we make use of the eigenvalue decomposition of the lattice nodel variable, i.e., of the sandpile model itself. This consists in finding a basis of decomposition for states of the sandpile models. Any sandpile state can be decomposed on a basis of $N_x \times N_y$ (here $48 \times 48$) eigenfunctions. Here we automatically generate such a basis by diagonalizing the covariance matrix of our model (see \citealt{strugarek_sandpile_2017} for more details), and we limit the number of eigenvalues to assimilate to 50. This effectively greatly reduces the dimension of our minimization while maintaining our ability to minimize adequately our cost function. The details of this technique can be found in \cite{hung_variational_2017} in the context of the prediction of the solar cycle, and in \cite{strugarek_sandpile_2017} in the context of flare forecasting with sandpile models.

The minimization of the cost function can be realized through standard down-gradient algorithms; advanced 4D-Var method using an adjoint code to estimate the gradient of the cost function \citep{belanger_predicting_2007} (and hence significantly reducing the convergence time of the minimization); or more generic and costly algorithm such as simulated annealing to avoid local minima traps by allowing up-gradient exploration. We tested each of these methods. The adjoint of the sandpile model was automatically generated using the \href{http://tapenade.inria.fr:8080/tapenade/index.jsp}{Tapenade} software \citep{pascual_extension_2006}. We used the simulated annealing algorithm implemented in \href{http://numerical.recipes}{Numerical Recipes} \citep{press_numerical_1992}. We found by employing these different approaches that only the simulated annealing algorithm was able to robustly succeed. Indeed, the cost function (\ref{eq:CostFun}) presents numerous local minima, and most down-gradient minimization methods are not able to find a sufficiently good minimum. An alternative approach could make use of neural networks to minimize $\mathcal{J}$, which we leave for future work.

The simulated annealing method allows occasional up-gradient explorations with a probability that depends on a parameter $T_s$ that represents the temperature, in analogy to annealing in metallurgy. In this work, $T_s$ is initialized at 10, and is then decreased over 19 iterations until it reaches $5\times 10^{-2}$. As proposed in \href{http://numerical.recipes}{Numerical Recipes} \citep{press_numerical_1992}, each iteration of the simulated annealing operates on a downhill simplex method. Here we use 20 iterations of downhill simplex for each simulated annealing iteration. The choice of the cooling schedule of $T_s$ and the number of iterations for both algorithms have been found empirically to achieve satisfying results while maintaining relatively high calculation speed.

\subsection{Quality of the data assimilation procedure}
\label{sec:assimilSynthetic}

\begin{figure}
    \centering
    \includegraphics[width=1\linewidth]{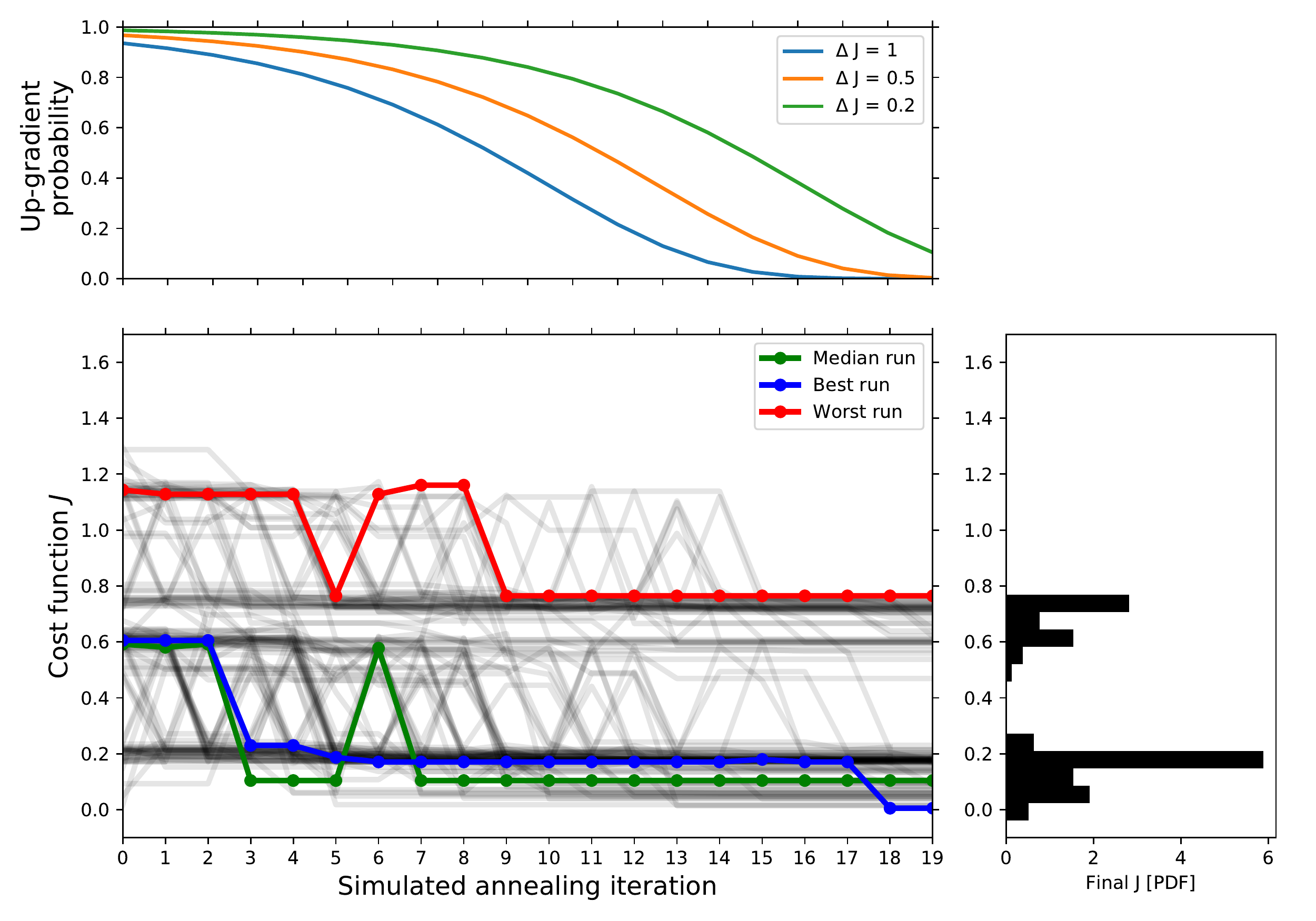}
     \centering
    \caption[Evolution of the cost function during 100 optimization runs using Model D01 on a single synthetic observation.]{Evolution of the cost function during 100 optimization runs of Model D01 on a single synthetic observation. The grey lines represent the results of all 100 simulated annealing runs, and the blue, green and red lines highlights the best, median and worst run respectively. The up-gradient probability of the algorithm is shown in the upper panel on the same timescale. The histogram on the right shows the distribution of final cost function values for the 100 runs.}
        \label{JvsIter}
\end{figure}

Figure \ref{JvsIter} highlights three examples of the evolution of the cost function $\mathcal{J}$ for Model D01 undergoing our minimization algorithm presented above. The upper panel traces the evolution of the up-gradient step probability, and the lower panel the evolution of the cost function. In the lower panel, 100 different simulated annealing runs are presented as grey lines. The median, worst and best runs are respectively shown in green, red and blue. In the initial phase of the minimization, the temperature $T_s$ is high enough for up-gradient exploration to occur fairly frequently, as seen in the upper panel. This can be seen in the initial phase of the worst (red) and median (green) runs, with a clear up-gradient step taken at the fifth simulated annealing step. As the temperature diminishes, the cost function stabilizes to a local minimum. As can be seen on the cost function histogram plotted at bottom right, the end result of these 100 minimization results contains two main groups, one with higher final cost functions of $\mathcal{J} \sim 0.5-0.7$ and the other with more successful runs of $\mathcal{J} \sim 0-0.2$. Such grouping of results are quite frequent in our minimization outputs, and result in part from the discrete nature of the penalties produced by misses and false alarms in the design of the cost function $\mathcal{J}$.

\begin{figure}
    \centering
    \includegraphics[width=0.45\linewidth]{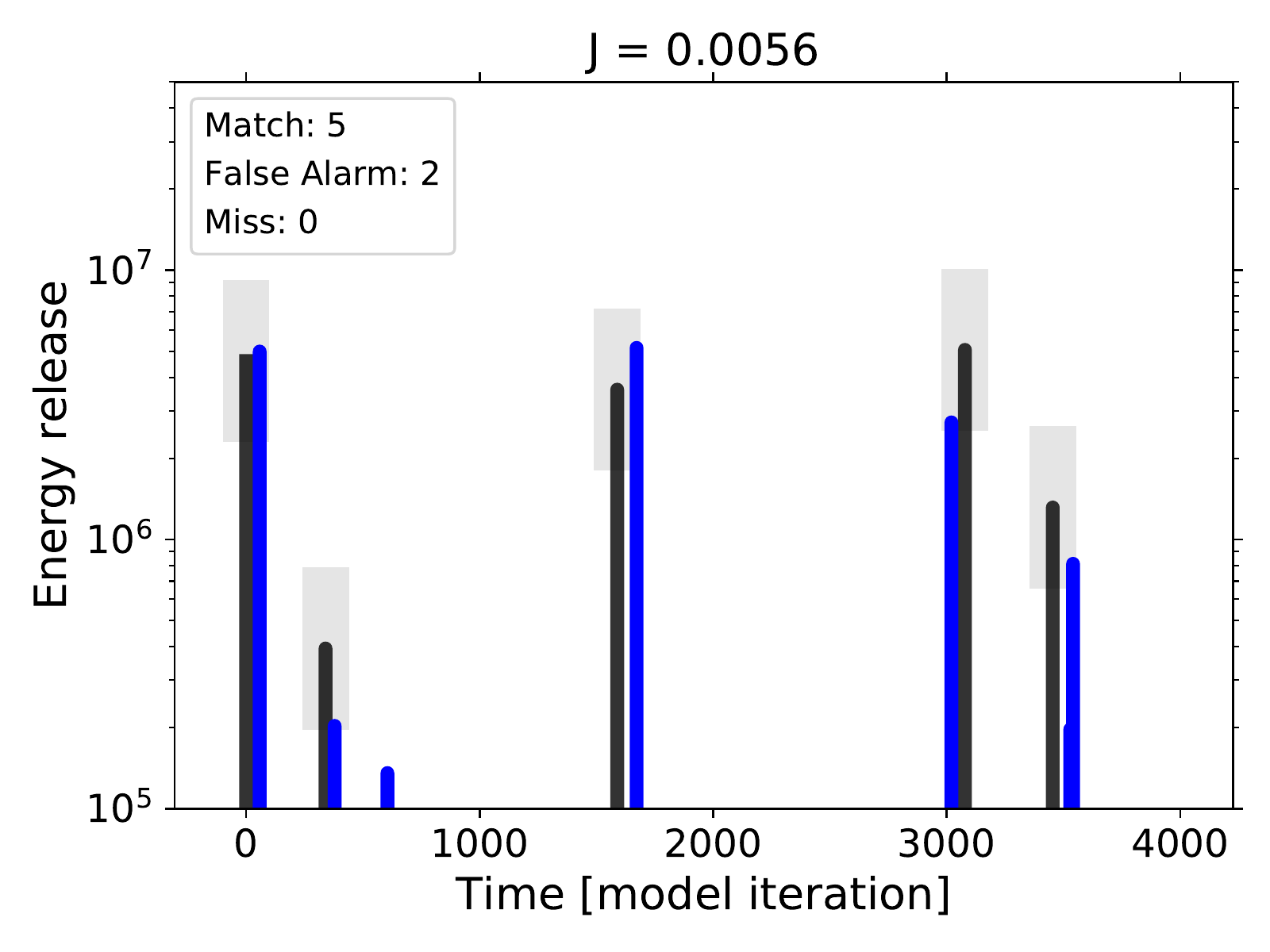}
    \includegraphics[width=0.45\linewidth]{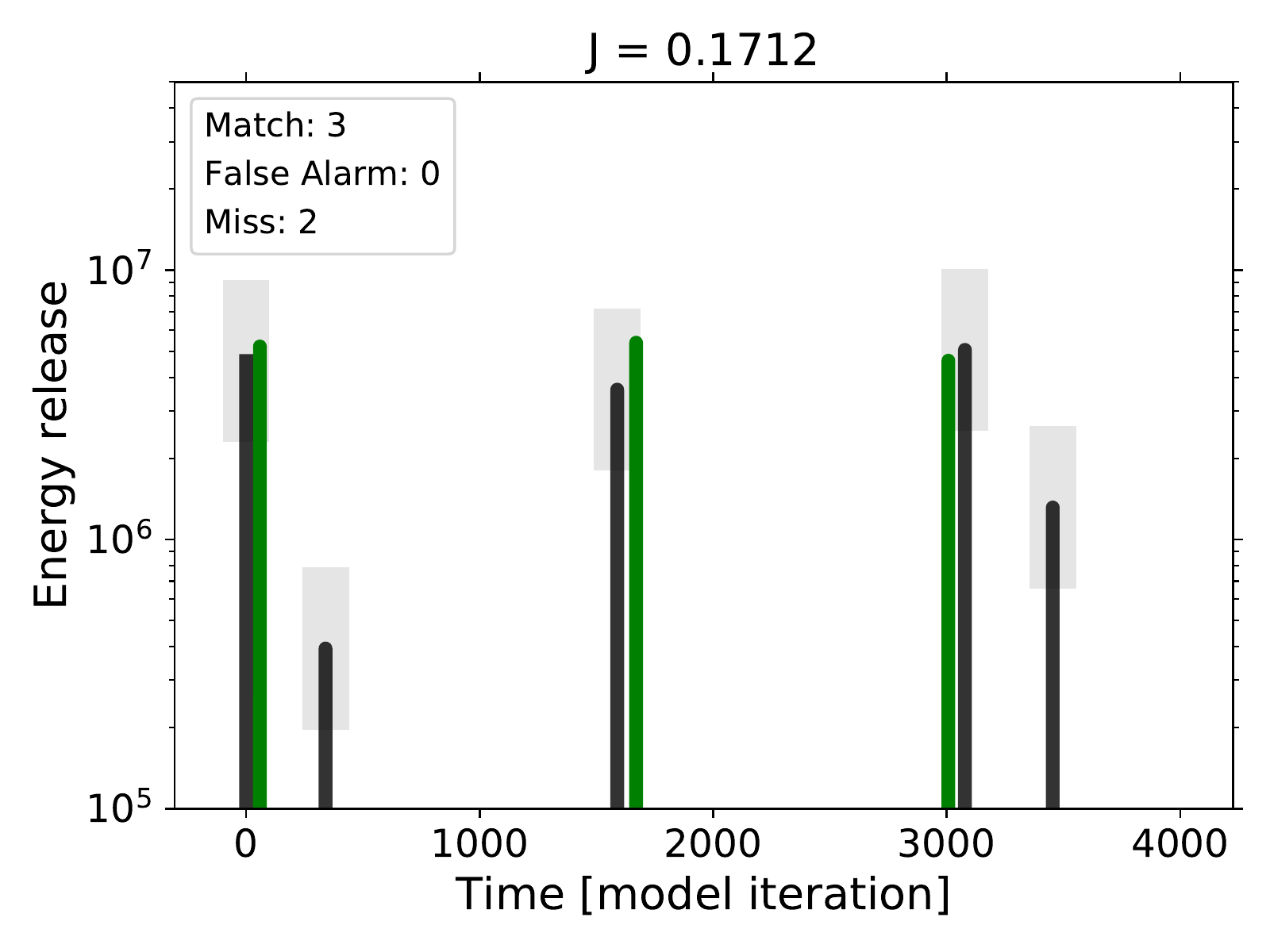}
    \includegraphics[width=0.45\linewidth]{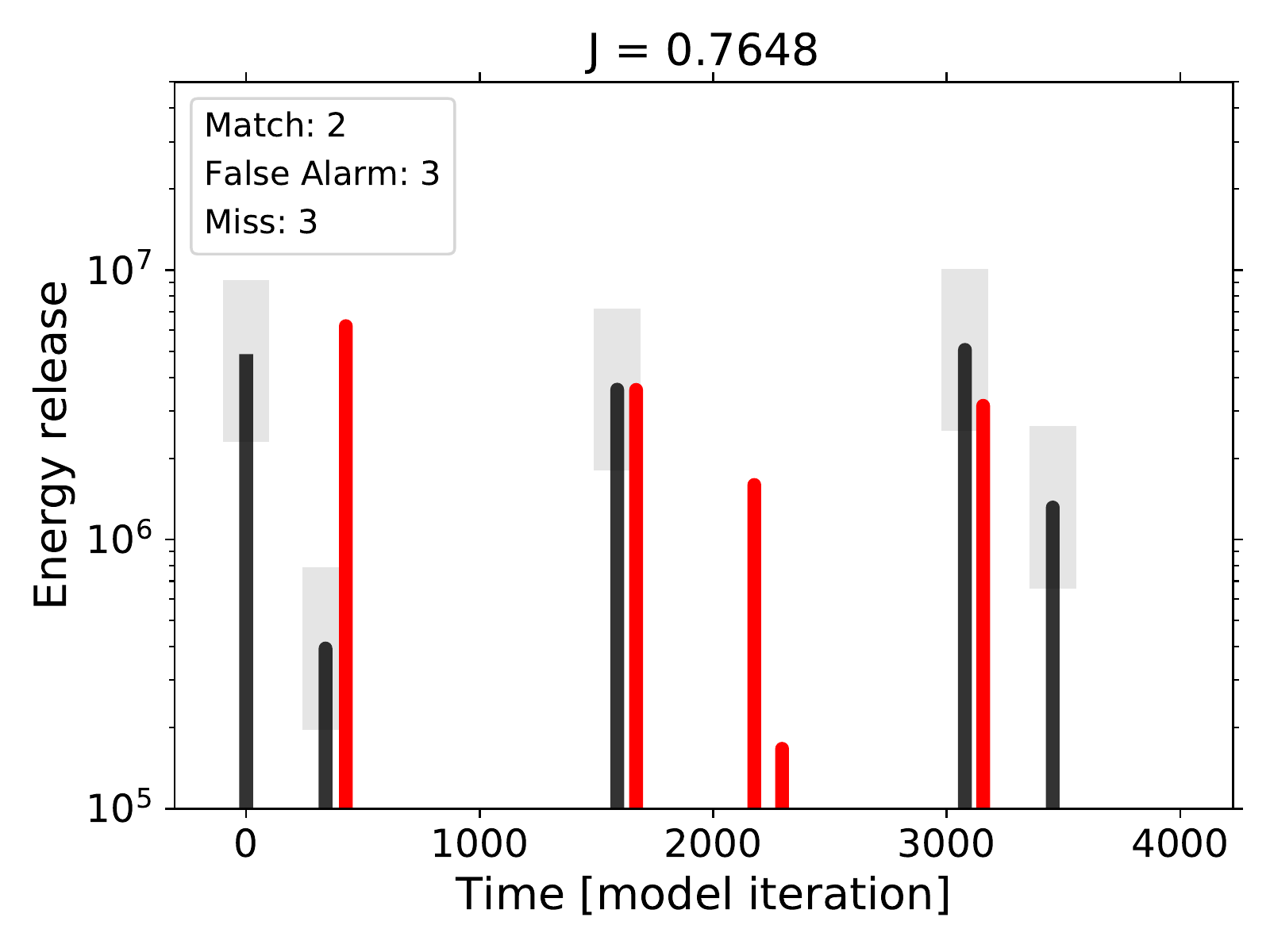}
     \centering
    \caption[Event energy time series of the best (left, blue), median (middle, green) and worst (right, red) output of 100 data assimilation runs using Model D01 on a common synthetic data.]{Event energy time series of the best (top left, blue), median (top right, green) and worst (right, red) output of 100 data assimilation runs using Model D01 on a common synthetic data. These three examples corresponds to the output of the same color-highlighted examples shown in Figure \ref{JvsIter}.}
        \label{DAExamples}
\end{figure}

We show in Figure \ref{DAExamples} the resulting compressed time series of avalanche energy release at the end of data assimilation. The color-code of Fig. \ref{JvsIter} has been retained and we illustrate the models resulting from the best (blue, top left), median (green, top right) and worst (red, bottom) assimilation runs.  In the first example, the assimilated run (blue vertical line segments) succeeds in capturing the five largest synthetic reference events (black), leaving as ``false alarms'' only two small events barely above our $10^5$ threshold (around iteration 600, and around iteration 3500 right before the last matched event in the window). This leads to a near-perfect cost function of $J = 0.0056$. In the second example (green peaks), two low-energy reference events are missed which leads to a cost function of $J=0.17$. Finally, in the third example (red peaks), only two reference events were captured, and three were missed. These large missed events lead to a large increase in our cost function, here $J=0.76$, as per the high penalty assigned to misses in our cost function design (viz.~\S\ref{S-cost}). 
% is mostly due to the three missed events (recall that our cost function is designed to have a bigger penalty on larger missed events). 
The three false alarms also contribute, to a lesser extent, to this higher cost function (misses have twice the weight as false alarms).  These three examples show variations of assimilation results on a single synthetic observation, but the results also fairly depend on the synthetic data being assimilated.

We now characterize the average performance of our data assimilation process. We produce randomly 100 synthetic observation timeseries, which we attempt to assimilate. For each synthetic observation, we run 100 assimilation runs. The results are summarized in Table \ref{tbl:ResultsDA_Jhist}. In the case of model D01 we find that the assimilated runs produce a satisfying cost function $\mathcal{J}\leq 0.1$ for 6\% of the assimilation runs. On average, the assimilated run rather give results with a cost function between 0.2 and 1. Conversely, if one randomly generates sequences of the avalanche models instead of using the assimilated runs, the average cost function increases significantly, and a satisfying cost function $\mathcal{J}\leq 0.1$ is achieved for only 1\% of the random sample. Our assimilated runs therefore clearly show significantly better results than the randomly generated initial conditions. Most importantly, the low computational cost of assimilating these sandpile models allows us to easily repeat the assimilation process until
an acceptable ($\mathcal{J}\leq 0.1$) assimilation is obtained, to later be used for
forecasting.
%times, only to keep the best runs for one reference data within the 6\% acceptance rate.     
%The top panel of Figure \ref{DAHist} shows a more general trend of our data assimilation results, produced by a set of 100 synthetic data, each being assimilated 100 times, for a total of 10 000 cost functions for each model D01 and D09 (labelled as 'assimilated'). In addition to the assimilated results, we included the cost function distributions of 10 000 non-assimilated initial conditions, again with 100 runs for each of the 100 synthetic data (labelled as 'random'). Each of these non-assimilated initial conditions was selected at random iterations of a simulation run of $\sim 10^8$ iterations. 

%Our assimilated runs clearly show significantly better results than the randomly generated initial conditions. They both exhibit similar results, a rising phase for low cost function, a peak at $J \sim 0.5$ with finally a diminishing tail until $J \sim 1.1 - 1.2$. The low consistency of our assimilation results is mainly due to the stochastic nature of our simulated annealing method and the variety of synthetic observations to fit. That being said, the low computational cost of assimilating these sandpile models leads no problem to repeat the assimilation process many times, only to keep the best runs for one reference data.

\begin{figure}
    \centering
    \includegraphics[width=0.75\linewidth]{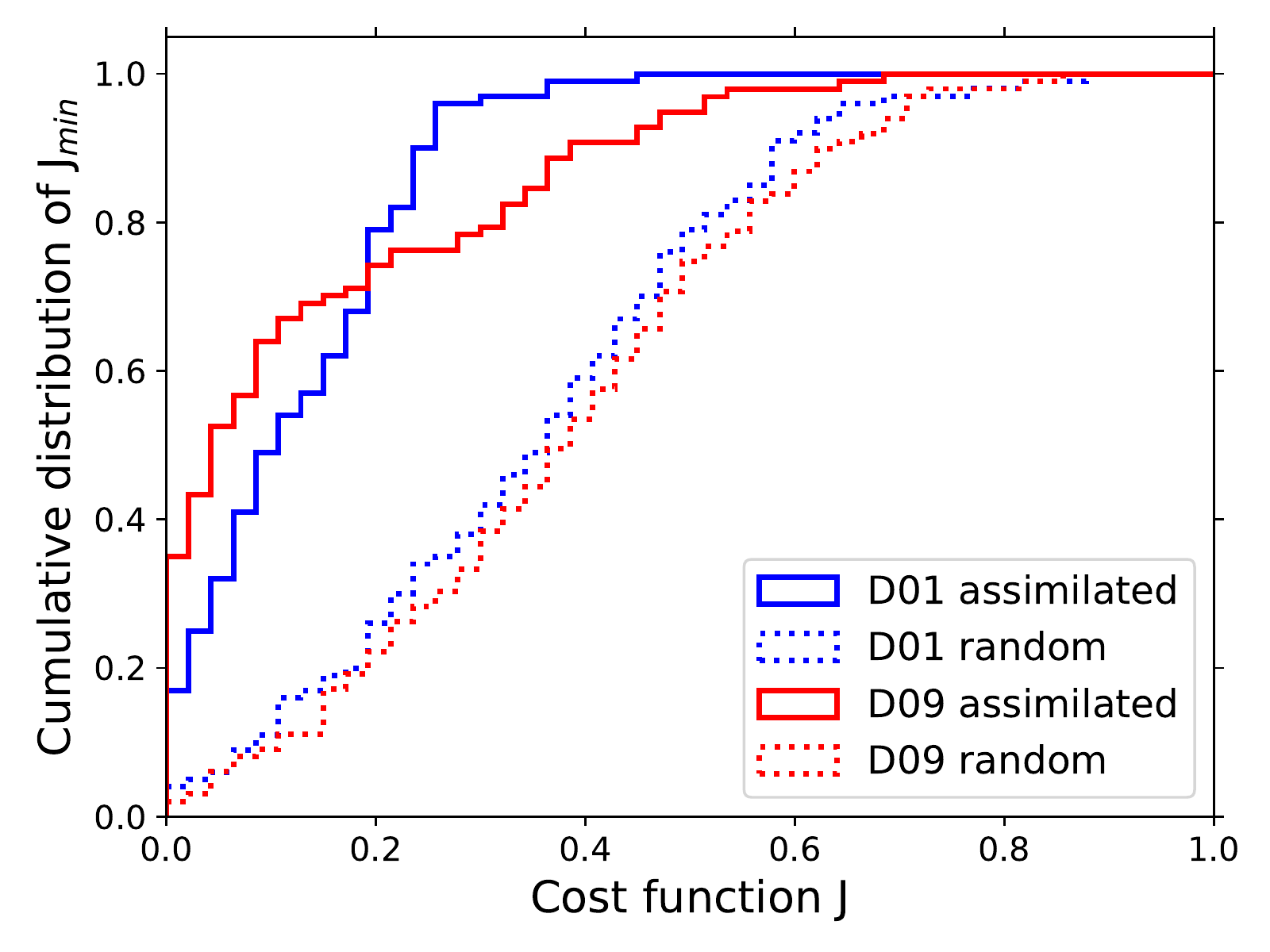}
     \centering
    \caption[Cumulative distribution of the best results ($J_{min}$) on each of the 100 synthetic observations.]{Cumulative distribution of the best results ($J_{min}$) on each of the 100 synthetic observations. The dashed lines represent the results generated from random states, whereas the full lines are from assimilated states. }
        \label{DAHist}
\end{figure}

We further characterize the data assimilation performance by the cumulative distribution of the best cost function achieved for each observation, denoted as $J_{min}$, and shown in Figure \ref{DAHist} for models D01 and D09. 
%The bottom panel of Figure \ref{DAHist} shows the cumulative distribution of the best cost function achieved for each observation, denoted as $J_{min}$. 
Both these models clearly outperform unassimilated runs from a random initial condition (shown as dotted lines). Only around 17\% of observations produce a $J_{min}$ close to 0 for model D01, and 36\% for model D09. A cost function of 0 is difficult to achieve for most observations with our current data assimilation algorithm, especially for those with the most events to assimilate. 
%It is probable that more allowed iterations during our simulated annealing method would help get better results. 
Model D09 outperforms Model D01 until around $J_{min} = 0.21$, but model D01 seems more robust with its best results, with all of them being at least under $J_{min} = 0.5$.

%In summary, Model D09 seems better suited to achieve at least some great results ($J_{min}<0.1$), but Model D01 produces more consistent results. 

\begin{table}
\caption{Results of all 10000 data assimilation runs ([\%])}\label{tbl:ResultsDA_Jhist}
\begin{tabular}{ccccc}     
J range  & D01 Assimilated &D09 Assimilated& D01 Random & D09 Random \\ 
 \hline
 $[0,0.1]$ &  6.48 &   5.90     &1.15  & 0.10 \\
 $[0.1,0.2]$ &  9.99  &  7.65  &0.36  & 0.17 \\
 $[0.2,0.5]$ &  40.67 &   43.17 &2.18  & 1.39 \\
 $[0.5,1]$ &  42.15  &   43.00  &15.38 & 11.15 \\
 $> 1$ &  0.71  &   0.25         &80.93 & 87.18 \\
 \hline
\end{tabular}
\end{table}

In all cases, the assimilated runs clearly outperforms randomly generated sequences. Even though the above results suggest that model D09 outperforms to some extent model D01 in yielding very low values for the cost functions, we also recall from \S\ref{sec:PrefCapAvalModel} that model D01 has a much longer prediction horizon (cf.~middle and bottom panels of Fig.~\ref{EventTimingDistribution}).
Consequently, all forecasting experiments reported upon in the remainder of this paper make use of model D01. 

\subsection{Predictive capabilities on synthetic data}
\label{sec:predSynthetic}

Before turning to real solar flaring data we first verify that the data assimilation process described above increases the forecasting skill over forecasts performed using a random (unassimilated) initial condition. The latter captures the model ``climatology'', i.e., the mean statistical behavior in the triggering
of avalanches. 

We adopt the ``all-clear'' forecasting paradigm described in \citet{2008ApJ...688L.107B} (see also \citealt{barnes_comparison_2016}). This is a binary forecast which
is defined as follows: given a forecasting window $t\in [0,T]$ extending from the present ($t=0$) to
some future time ($t=T$), will at least one avalanche/flare of energy $E>E^\ast$ occur at any time within this window. %In what follows we set its temporal width $T$ to twice the median inter-event waiting time for an extended run of model D01, $T=1138\,$iterations.
We measure forecasting performance through a simple ``Rate Correct'' (RC) measures:

\begin{equation}
\label{eq:RC}
    {\rm RC}={{\rm TP}+{\rm TN}\over N}~.
\end{equation}
where TP is the count of true positives (an avalanche was predicted in the forecasting window and did occur), TN the count of true negatives (no avalanche was predicted and none occurred),
and $N$ is the total number of forecasts attempts. 
We extract 250 non-overlapping 4000-iteration long segment of energy release compressed time series from an extended simulation run of model D01. We then use the first 2000 iterations of each as the assimilation window, and the subsequent portion of the segments as target. For each such dataset, we run our data assimilation scheme 100 times using 100 distinct random initial conditions, and group the resulting initial conditions in terms of the cost function values attained. For each member of each group for each synthetic dataset,
we then perform an ensemble of $10^4$ statistically-independent all-clear forecasts using the same assimilated initial condition but distinct random number streams controlling nodal variable losses during avalanches.  
The Rate Correct statistics presented below are thus based on a total of
anywhere from $10^4$ up to $N=10^6$ of such forecasts.

\begin{figure}
    \centering
    \includegraphics[width=0.99\linewidth]{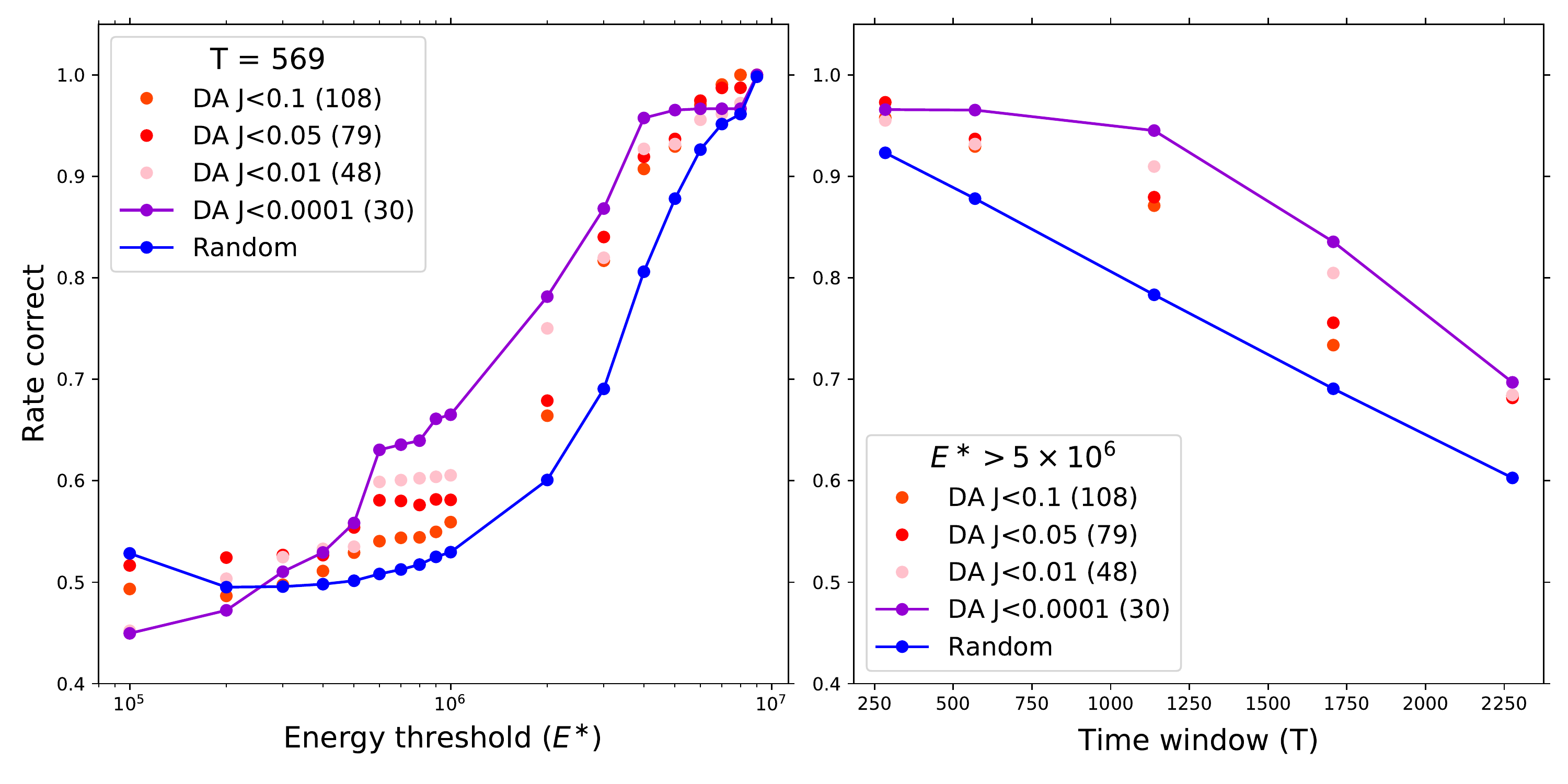}
    
     \centering
    \caption{``All clear'' forecasting performance on synthetic data. The left panels shows the variation of our rate correct measure (eq.~\ref{eq:RC}) as a function of target avalanche energy threshold,
    expressed in units of $e_0$ (viz.~eq.~\ref{Eq-e0}) over an ``All-Clear'' forecasting window of $T=569$ model iterations. The right panel shows an equivalent plot, but now as a function forecasting window for a fixed target avalanche energy threshold of $E^\ast>5\times 10^6$. The color scheme identifies the range of cost function values used to select assimilated initial conditions for forecasting, as labeled, with the corresponding number of members in each ensemble given in parentheses. The blue dots and line refer to forecasts produced from a random initial condition, which defines our model climatology (see text).}
        \label{RateCorrectSynthetic}
\end{figure}

The left panel of Figure \ref{RateCorrectSynthetic} displays the variations of the Rate Correct
measure as function of the threshold (energy lower limit) values $E^\ast$ (measured in units of $e_0$, viz.~eq.~\ref{Eq-e0}) imposed to observations within the all-clear forecast window. The latter is here of a fixed duration of $T=569$ model iterations, corresponding to the mean wait time
for avalanches having energies larger than $E^\ast=10^5$. The blue dots show the model climatology, constructed by using an unassimilated initial condition as input. Other colors indicate forecasting performance
for ensembles of assimilation runs increasingly stringent on the cost function value deemed acceptable to be included in the forecast ensemble, as labeled. 
At low $E^\ast$ all ensembles yield a rate correct of 0.5, consistent with the median waiting time, and consistent with the notion that our forecasting scheme is not expected to perform well for small avalanches, where stochasticity dominates the avalanching behavior (viz.~Fig.~\ref{EventTimingDistribution}). At very high $E^\ast\simeq 10^7$, avalanches are very rare in both target and forecast time series, so all forecast inevitably converge to RC$\simeq 1$. However, for the large avalanches in the approximate range
$10^6\leq E^\ast \leq 5\times 10^6$, assimilated initial conditions with cost function $J<0.0001$ improve the all-clear forecasting performance
by some 20\% to 30\%

The right panel of Figure \ref{RateCorrectSynthetic} displays the results of the same ensemble of forecasting experiments, but now as a function of the extent of the forecast window ($T$), at fixed threshold energy $E^\ast=5\times 10^6$. At very small $T$ the probability of producing a large avalanche is small, so all ensembles perform well (RC$\to 1$) by forecasting ``no avalanche''. At $T$ approaches the median waiting time for this threshold value of $E^\ast$, all ensembles tend towards
RC$\simeq 0.5$. In between, assimilated runs with $J<0.0001$ (purple) again outperform the climatological forecast (random, in blue) by up to 30\%.

\section{Forecasting large flares from GOES X-Ray flux time series}
\label{sec:GoesDA}

We now turn to the assimilation and prediction of solar flare data. We use again model D01 for all results presented in this section, in views of its superior predictivity level demonstrated in \S\ref{sec:PrefCapAvalModel} (viz.~Fig.~\ref{EventTimingDistribution}). 

\begin{figure} 
\centerline{\includegraphics[width=0.48\linewidth]{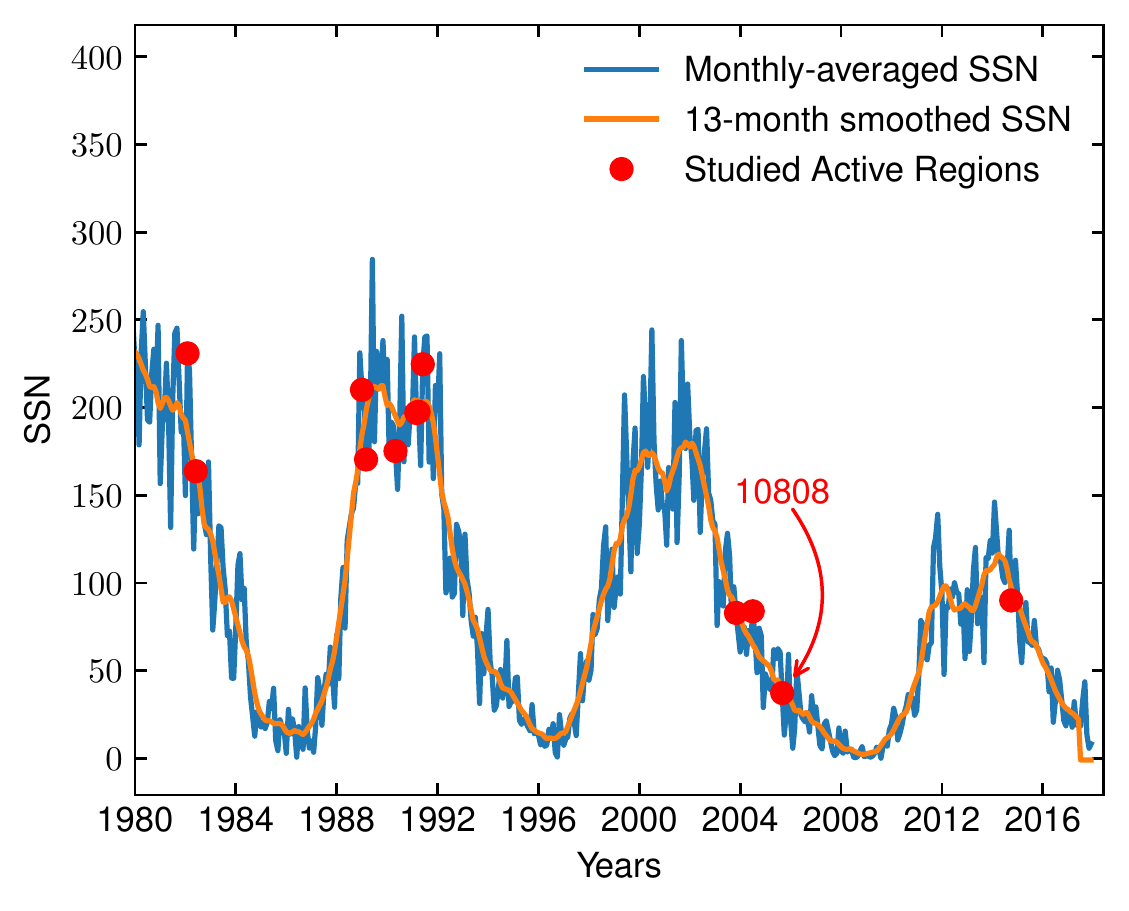}
\hfill\includegraphics[width=0.48\linewidth]{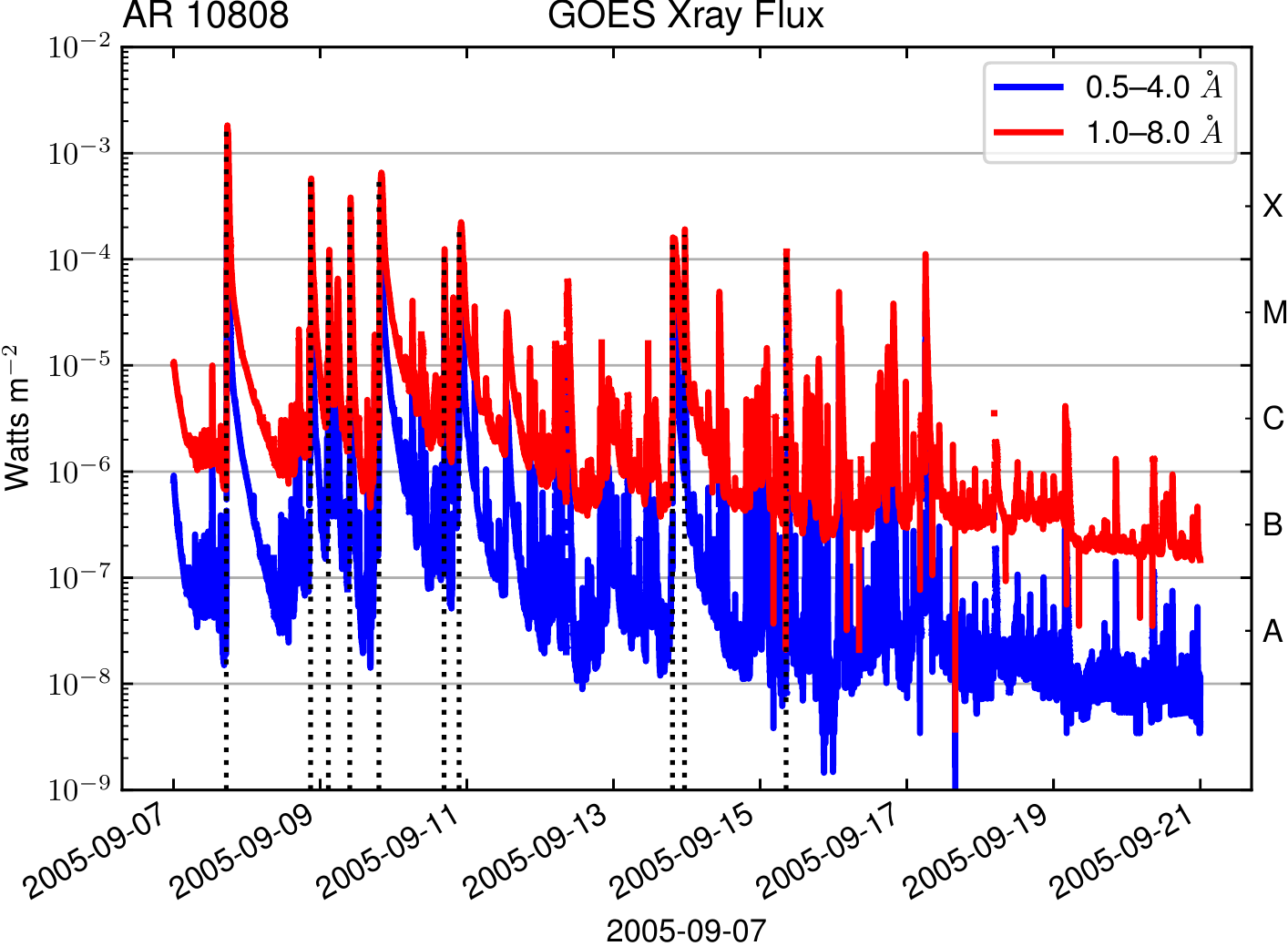}}
\caption{\textit{Left}: Monthly-averaged (blue) and 13-month smoothed (orange) sunspot
  number (SSN) of the past 36 years (data from
  \href{http://sidc.oma.be/silso/home}{SILSO}, Royal Observatory of
  Belgium, Brussels). The identified active regions reported in Table
  \ref{tbl:GOESandHEKdata} are labelled by the red
  circles. \textit{Right}: GOES time series for AR 10808, with the 6
  flares of class X identified in the Heliospheric Events
  Knowledgebase, indicated as vertical dotted lines. The two channels of the GOES XRS instrument are shown
in blue (0.5-4 {\AA}) and red (1-8 {\AA}).}\label{fig:ARandCycle}
\end{figure}

Our scheme requires information on past flaring behavior in order to forecast future flaring. We opted to secure such
 a discrete series of flaring events from the continuous monitoring of the Sun in X-ray by the Geostationary Operational Environmental Satellites (GOES). The sandpile model (Section \,\ref{sec:AvalModels}) can be conceptually related to the magnetic field overlying a single active region. The GOES timeseries, on the other hand, are obtained from the full solar disk and thus intertwine flaring events from different active regions. As a first selection step, we identify past epochs of solar activity when large flares (at least four flares of GOES class X and above) were triggered by only one predominant active region of the Sun. %\textcolor{red}{{\bf phrase suivante dans conclu? distraction ici}An assembly of active regions could in principle be represented by a series of sandpile models, with the possibility to couple them to take into account sympathetic phenomena. This aspect is left for future work and we focus here on past individual flaring active regions.}  \AS{Ok for me to put that in conclusions}
%As we base our approach on using the past flaring history of an active region, 
%We further limit our sample to active region which produced their first X-class flare at a Carrington longitude larger than \todo{X}, to ensure the history of the active region was indeed captured in the GOES observations. 
Based on these criteria we identified 11 past epochs with strong flaring activity arising from a single, large active region. These are listed in Table 
\ref{tbl:GOESandHEKdata}, and sample various phases of the solar activity cycle, as shown
on Fig.~\ref{fig:ARandCycle}.

For each epoch, a series of event can be constructed by querying the Heliospheric Events Knowledgebase (HEK), or by identifying flares from the GOES X-ray flux in the 1-8{\AA}  band following the algorithm detailed in \citet{aschwanden_automated_2012}. An example of the two types of series of events is show in Fig. \ref{fig:ARandCycle} for the active region AR 10808. In the following we will only use the time series constructed from the HEK, as it allows to select only flares produced by a given active region. The corresponding list of events (time and peak flux), plotted as vertical dotted lines on Fig.~\ref{fig:ARandCycle} is treated as the observational counterpart to the avalanche model's compressed time series for energy release (viz.~Fig.~\ref{EnergyReleaseCompression}).

%\begin{table}
%\caption{Selected active regions}\label{tbl:GOESandHEKdata}
%\begin{tabular}{llccccc}     
%AR  & Time period & Maximum Flux & \multicolumn{3}{c}{Number of Flares} & \# of other
%  ARs \\ % & Simultaneous active regions \\
%number & & [$10^{-4}$ W/m$^2$] & X & M & C & during the same period \\ % &
%                         % producing (\#) X-class flares \\ 
% \hline
%  6555 &  1991/03/17 - 1991/03/31 &   9.4 &  7 &  28 &  44 &               8 \\
%  3576 &  1982/01/26 - 1982/02/09 &   2.6 &  6 &  14 &  32 &              14 \\
% 12192 &  2014/10/16 - 2014/11/01 &   3.1 &  6 &  32 &  74 &               2 \\
%  5747 &  1989/10/14 - 1989/10/27 &  13.0 &  5 &  21 &  27 &               6 \\
% 10720 &  2005/01/11 - 2005/01/23 &   7.1 &  5 &  17 &  66 &               3 \\
%  5629 &  1989/08/03 - 1989/08/17 &  20.0 &  5 &  17 &  32 &               6 \\
%  5047 &  1988/06/15 - 1988/06/27 &   5.6 &  4 &   6 &  28 &               2 \\
% 11748 &  2013/05/12 - 2013/05/21 &   3.2 &  4 &   5 &  19 &               3 \\
% 12673 &  2017/09/03 - 2017/09/11 &   9.3 &  4 &  26 &  54 &               2 \\
%  4026 &  1982/12/11 - 1982/12/23 &  12.0 &  4 &  12 &  22 &               5 \\
% \hline
%\end{tabular}
%\end{table}

\begin{table}
\caption{Selected observed active regions.}\label{tbl:GOESandHEKdata}
\begin{tabular}{llcccc}     
AR  & Time period & Maximum Flux & \multicolumn{3}{c}{Number of Flares} \\ % &  Simultaneous active regions \\
number & & [$10^{-4}$ W/m$^2$] & X & M & C \\ % & (\# X-class flares) \\ 
 \hline
 %3576 &  1982/01/26 - 1982/02/09 &   2.6 &   6 &  14 &  32 &                       \\
 %3763 &  1982/06/02 - 1982/06/15 &  12.0 &   6 &  52 &  37 & \\ %             3776 (2) \\
 5312 &  1989/01/06 - 1989/01/20 &   2.3 &   6 &  33 &  12 \\ % &                       \\
 5395 &  1989/03/05 - 1989/03/19 &  15.0 &  11 &  48 &  48 \\ % &                       \\
 6063 &  1990/05/11 - 1990/05/24 &   9.3 &   5 &   8 &  10 \\ % &                       \\
 6538 &  1991/03/05 - 1991/03/17 &   5.5 &   5 &  17 &  38 \\ % &    6537 (1), 6545 (5) \\
 6545 &  1991/03/11 - 1991/03/22 &   3.9 &   6 &  16 &  35 \\ % &             6555 (1) \\
 6555 &  1991/03/17 - 1991/03/31 &   9.4 &   7 &  28 &  44 \\ % &              6545 (1) \\
 6659 &  1991/06/01 - 1991/06/17 &  12.0 &   6 &  28 &  38 \\ % &                       \\
10486 &  2003/10/22 - 2003/11/05 &  28.0 &   7 &  20 &  16 \\ % &  10484 (1), 10488 (2) \\
10649 &  2004/07/12 - 2004/07/24 &   3.6 &   6 &  10 &  46 \\ % &                       \\
10808 &  2005/09/07 - 2005/09/19 &  17.0 &  10 &  20 &  47 \\ % &                       \\
12192 &  2014/10/16 - 2014/11/01 &   3.1 &   6 &  32 &  74 \\ % &                       \\
 \hline
\end{tabular}
\end{table}

Our next task is to match the temporal and flux scale of GOES observations to their equivalent in the avalanche model. We first threshold the 1-8\AA\ GOES time series to retain only flares of class X, and establish a linear scaling between peak flux measured by GOES to (compressed) peak energy from the model, matching the most intense GOES flare in Table 2 to $5\times 10^6$, the largest avalanche energy
produced by a typical extended model run.
%At least in avalanche models, the peak energy release does scale close to
%linearly with total energy release (see, e.g., Fig.~7 in Lu et al.
Next we convert the GOES time stamps to model iteration by requiring that the median inter-event wait time matches approximately the median wait time of the model. This is done by multiplying the GOES time stamps by a conversion factor, chosen to ensure that the resulting time series segments contains no less than 6 but no more than 9 X-class flares per 4000 model iterations, the width of our assimilation window in all results presented below.
We carry out this manual adjustment separately for each active region in Table \ref{tbl:GOESandHEKdata}.
%We ensure that the resulting time series segments contains no less than 6 but no more than 9 X-class flares per 4000 model iterations, the width of our assimilation window in all results presented below.  
This procedure, as minimalistic as it might be, at least captures intrinsic variations in flaring rates and peak energy release from one active region to another.

\begin{figure}
    \centering
    \includegraphics[width=1\linewidth]{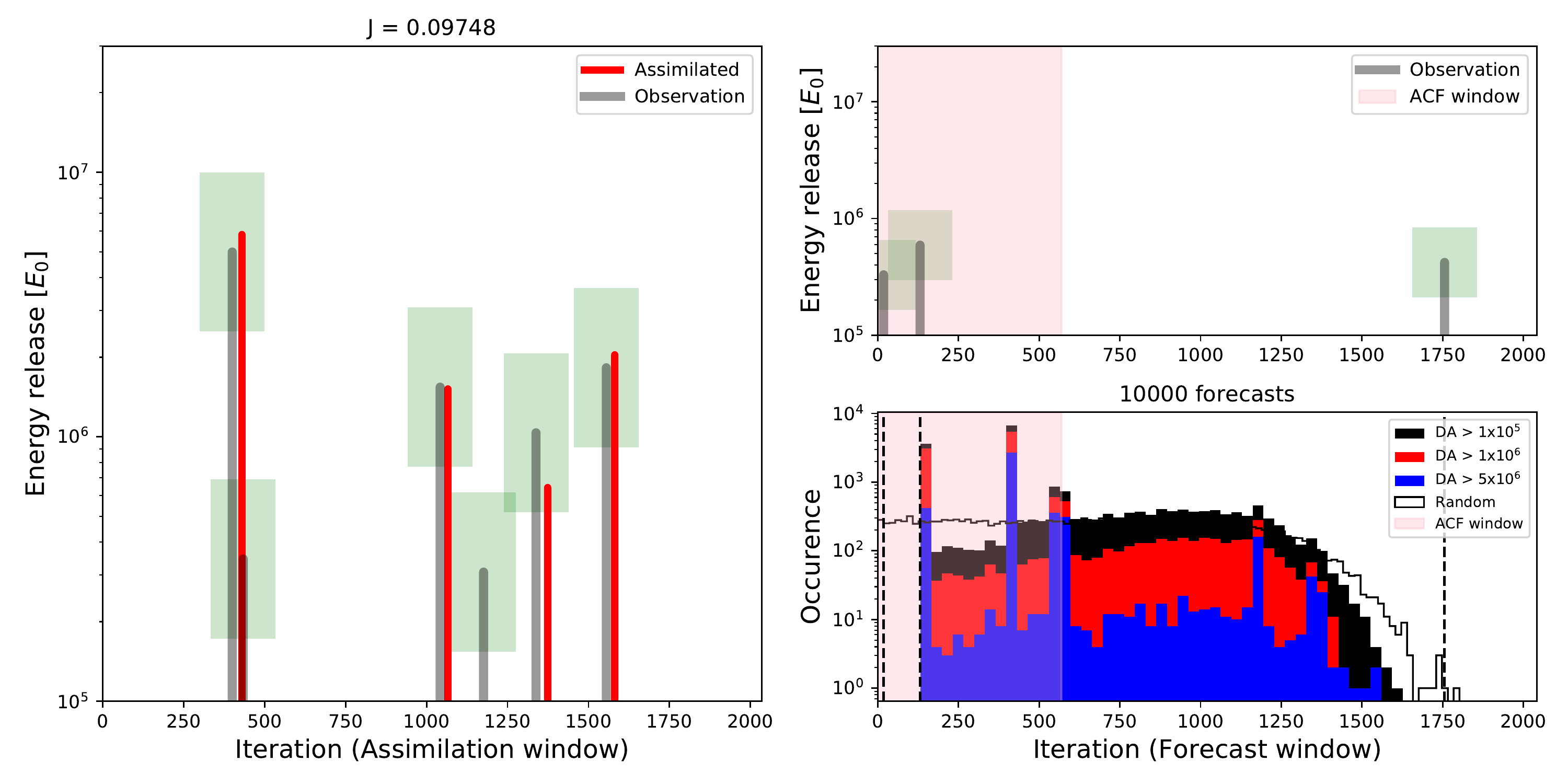}
     \centering
    \caption[Description.]{\textit{Left}: Assimilation window results for a representative assimilation run using Model D01 (red) over GOES AR number 10808 compressed time series (grey). \textit{Top right}: Forecast window for the GOES AR number 10808 time series. \textit{Bottom right}: Composite temporal distribution of events in all 10000 forecasts produced from Model D01 following the assimilation window, varying only the random number seed for each forecast. The dash vertical lines represent the timing of the target observation. The pink area spans the width of the "All clear forecast" window.}
        \label{Forecast_GOES_Example}
\end{figure}

Figure \ref{Forecast_GOES_Example} shows a representative example of the assimilation of GOES data for AR10808 and subsequent ``all clear'' forecast.
Here the assimilation (left panel) properly captured 5 of 6 events in the assimilation, missing only the lowest energy event at iteration 1150, for a final cost function value of $\mathcal{J}=0.0975$. The top right panel shows the temporal continuation of the GOES data, with our all-clear forecast window indicated in pink. The bottom right panel shows the superposition of 10000 forecasts produced using the assimilated initial condition. The three occurrence
histograms shown (note the logarithmic vertical scale) are color coded according to total avalanche energy, as labeled. As in Figure \ref{EventTimingDistribution}, the presence of peaks indicated flaring behavior occurring at preferred times. During the first 125 iterations of the forecast model, no strong avalanche is produced, which is why the histograms are empty. Then 97.37\% of forecasts do produce at least one avalanche within the forecast window. In contrast, the model climatological forecasts, using 10000 unassimilated initial conditions (thin line histogram), has a significantly lower success rate of 61.89\%.
% DONE \textcolor{red}{Christian: insert the percentage success here}

%In discussing this Fig we need to remind the reader that the time of occurrence of the first flare
%is set by initial condition and driving rate, therefore for assimilated runs it always the same; but not the energy of that first flare because once avalanche starts stochasticity kicks in. We should have exolained this before in discussing Fig 4, but a bit repetition never hurts, and here is actually needed I think

\begin{figure}
    \centering

    \includegraphics[width=0.99\linewidth]{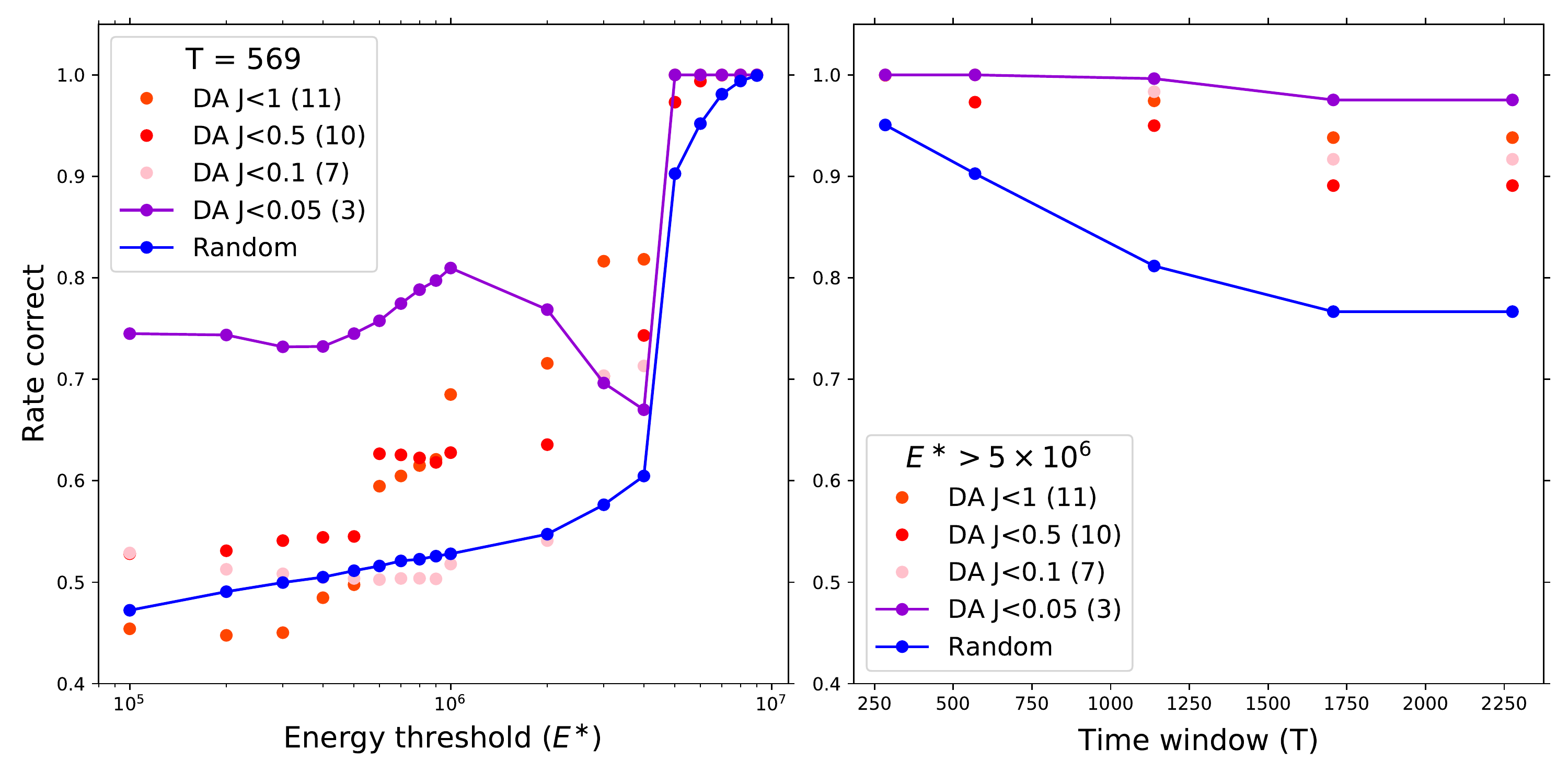}
    
     \centering
    \caption{``All clear'' forecasting performance on GOES data. The left panels shows the variation of our rate correct measure (eq.~\ref{eq:RC}) as a function of target avalanche energy threshold over a forecasting window of $T=569$ model iterations.
    The right panel shows an equivalent plot, but now as a function forecasting window a for fixed target avalanche energy threshold of $E^\ast=5\times 10^6$. The color scheme identifies the range of cost function values used to select assimilated initial conditions for forecasting, as labeled, with the corresponding number of members in each ensemble given in parentheses. As on Fig.~\ref{RateCorrectSynthetic}, the blue dots and line give the model's climatological forecast.}
        \label{RCvsThreshol_GOES}
\end{figure}

Figure \ref{RCvsThreshol_GOES}, similar in format to Fig.~\ref{RateCorrectSynthetic}, shows the true correct rate defined by eq.~\ref{eq:RC}, against increasing threshold $E^\ast$ in the energy of the target GOES flares over a forecast window of $T=569$ iterations (left), and against forecast window
width $T$ at fixed threshold energy $E^\ast=5\times 10^6$ (right).
Results are shown again for decreasing cost function cutoffs in the ensemble of assimilation runs retained
in forecasting, as color-coded and labeled. For each assimilated initial condition retained in each ensemble for a given cost function cutoff value, 1000 forecasts are produced to generate the results displayed
in the Figure. The model climatological forecast
associated with 1000 forecasts for each one of 11 unassimilated initial conditions is again shown in blue.
As with the forecasting of synthetic data discussed in the preceding section, forecasts of GOES flaring data based on assimilation outperform model climatology at all energy threshold values, but the improvement is greatest for flares in the middle of the plotted (logarithmic) range. As discussed previously, decreasing forecasting skills for small avalanches/flares is expected, as these are more strongly affected by the stochastic elements present in the model, with the associated lattice stress patterns not necessarily well represented by an eigenvalue decomposition retaining only the 50 highest-power modes. 
%A rapid drop in forecasting skill is also apparent for the very largest avalanches ($\simeq 10^7 e_0$).
%These rare events are characterized by a break of scale invariance, because avalanches
%reach to a large fraction of lattice boundaries, and are thus strongly constrained/influenced by the adopted lattice size and geometry.

%\begin{figure}
%    \centering
%    \includegraphics[width=0.75\linewidth]{img/article/RCvsThreshold_log.pdf}
%     \centering
%    \caption[Description.]{Rate correct skill score produced from 10000 forecasts on 11 GOES X-Ray flux time series, for a total of 100000 forecasts. In red represents forecast produced from assimilated state, in blue from random lattice states.}
%        \label{RCvsThreshol_GOES}
%\end{figure}

%\textcolor{red}{
%Do we want to also mention briefly D09 results here ? (to be clarified/decided)}

% meme figure que la figure 9, mais sur les données GOES

\section{Discussion and Conclusion\label{sec:conclusion}}

Lattice-based sandpile models of solar (and stellar) flares are a highly idealized representation of the processes of slow energy accumulation and rapid, instability-driven release in coronal magnetic fields structures, much along the lines of \citet{parker_nanoflares_1988}'s nanoflare model. In this paper we have described a solar flare forecasting method based on assimilation of GOES X-Ray flux time series into a sandpile model. The fundamental idea underlying this approach harks back to a known feature of self-organized critical avalanche models, namely the presence of long-range correlations in avalanching behavior. More specifically, large avalanches release large-scale stress patterns established across the lattice by earlier avalanches, in particular large ones. As a consequence, while the triggering and unfolding of small avalanche is dominated by stochastic effects and thus effectively unpredictable, this is not necessarily so for the larger avalanches. In section \ref{sec:PrefCapAvalModel} of this paper we demonstrated explicitly this somewhat counter-intuitive property of some self-organized critical sandpile models.

Distinct sandpile models can be constructed based upon specific choices of driving, stability threshold and redistribution rules. As demonstrated in \S\ref{sec:PrefCapAvalModel}, it turns out that such distinct model implementations exhibit widely varying levels of predictability even for the larger avalanches they generate. Of the three types of sandpile models tested, the more strongly non-conservative deterministically-driven models (model D01 herein) showed the longest predictability window. This in agreement with the earlier exploratory analysis presented in \citet{strugarek_predictive_2014}. Stochastically-driven conservative models performed the worst, consistent with the results presented in \citet{belanger_predicting_2007}. Most pertinent for its forecasting potential, the predictability window of the D01 model was found to extend over 4 times the median inter-event wait time characterizing energy release in large events.

Data assimilation in sandpile models requires adjusting the initial lattice state so as to reproduce avalanching/flaring behavior in a pre-set assimilation window where observations are available. This defines an optimization problem, which turns out to be a very challenging one. The form of cost function being minimized leads to a search space that is multimodal and strongly degenerate, as a great number of lattice configurations can lead to similar temporal avalanching patterns. We opted for an iterative optimization approach based on ensemble trials of simulated annealing, which allows adequate and consistent minimization of the cost function measuring misfit between model predictions and data being assimilated. The computational speed of the sandpile model is essential for achieving this minimization in a reasonable amount of computing time.

After validating our forecasting scheme on synthetic data in \S\ref{sec:predSynthetic},
we then applied our method to the ``all-clear'' forecast problem, working off GOES X-Ray flux time series for 11 solar active regions having generated multiple X-class flares (\S\ref{sec:GoesDA}). Our data assimilation scheme was shown to lead to a marked improvement in forecasting skill for the larger flares, as compared to the model's climatology. The latter was generated by simply skipping the data assimilation step, and running the forecasting step from an unassimilated initial condition randomly chosen from a reference run of the sandpile model. Such a forecast thus captures the mean ``flaring rate'' of the model, and as such represents an appropriate form of baseline climatology in this context. Forecasting improvement is greatest for the largest flares/avalanches, provided these are not large to the point of being affected by the break of scale invariance produced by the finite size of the lattice.

These results are very encouraging, but numerous improvements are possible, and indeed are needed. Most pressing perhaps is to improve on the very simple rescaling procedure introduced here to match the flaring rates and peak energy release inferred from GOES X-Ray time series to the corresponding quantities in the sandpile model. In all models considered here, a statistically uniform flaring rate results from a constant or statistically stationary driving rate, leading to an inter-event waiting time distribution of exponential form. This is not the case for flaring on the Sun, but the observed waiting time distribution can be reproduced reasonably well by partitioning the data into contiguous blocks of constant flaring rates. (e.g. \citealt{wheatland_origin_2000}). Such non-uniform flaring rates are readily generated in sandpile models by modulating the driving process, which can lead to waiting-time distributions better aligned with observations \citep{norman_waiting-time_2001}. Simultaneously matching model and observed two-dimensional distribution of events in waiting time and energy release is a promising avenue we are currently exploring. A proper scaling of energy and time between model and observation is also a prequisite to the calculation of a true skill score against observed flaring climatology \citep[for more on this point see][\S3]{barnes_comparison_2016}.

An important finding of the present study is the demonstration that not all self-organized critical sandpile models of solar flares have equal predictive potential, even though they may all match more or less equally well the observed power-law distributions of energy release in flares. Forecasting improvement may be achieved by identifying other sandpile models with even better predictive potential than our ``best'' model D01. Two candidates currently under examination are the fieldline-based model
of \citet{morales_self-organized_2008}, and the energy minimizing lattice model of \cite{farhang_principle_2018,farhang_energy_2019}.

There are inherent limitations to the flare forecasting scheme introduced here. Our data assimilation scheme requires a history of past flaring behavior for the active region being modeled, spanning a time interval of a few times the inverse flaring rate. This is quite different from many forecasting schemes that work off, say, line-of-sight or vector magnetograms of active regions, where in some cases a useful forecast can be produced from a single such magnetogram (see, e.g., the various forecasting schemes investigated in \citealt{leka_comparison_2019}). Our approach should be viewed as complementary to these existing techniques, especially since its low computational requirements could allow rapid,
near-real-time assessment of upcoming probable flaring activity of a given active region, to be then observed and assessed more closely with other methods.

\textbf{Acknowledgements} \ \ We thank A.S. Brun, L. Jouve  and A. Vincent for multiple discussions on data assimilation procedures. We acknowledge the financial support from the Natural Sciences and Engineering Research Council of Canada (NSERC). C.~Thibeault is supported in part by a merit scholarship from Hydro Qu\'ebec. A. Strugarek thanks N. Vilmer and L. Klein for discussions on solar flares and self-organized criticality models. A. Strugarek acknowledges funding from ERC WHOLESUN 810218 grant, INSU/PNST, CNES Solar Orbiter, and from the P2IO LabEx (ANR-10-LABX-0038)  in the framework "Investissements d'Avenir" (ANR-11-IDEX-0003-01) managed by the Agence Nationale de la Recherche (ANR, France). This work used data provided by the MEDOC data and operations centre (CNES / CNRS / Univ. Paris-Saclay), http://medoc.ias.u-psud.fr/.

%\acknowledgment US spelling: \verb+\acknowledgment+
%\acknowledgement British  spelling: \verb+\acknowledgement+

%%%%%%%%%%%%%%%%%%%%%%%%%%%%%%%%%%%%%%%%%%%%%%%%%%%%%%%%%%%%%%%%%%%%%%%%%%%
% \appendix   

     % format of references provided by the journal (.bst)
\bibliographystyle{spr-mp-sola}
     % name your Bibtex file containing your references (.bib)

\bibliography{Article_bib_18_05_2021.bib}

\begin{thebibliography}{36}
% BibTex style file: spr-mp-sola.bst (nameyear), 2015-03-09
\ifx\bisbn     \undefined \def\bisbn  #1{ISBN #1}\fi
\ifx\binits    \undefined \def\binits#1{#1}\fi
\ifx\bauthor   \undefined \def\bauthor#1{#1}\fi
\ifx\batitle   \undefined \def\batitle#1{#1}\fi
\ifx\bjtitle   \undefined \def\bjtitle#1{\textit{#1}}\fi
\ifx\bvolume   \undefined \def\bvolume#1{\textbf{#1}}\fi
\ifx\byear     \undefined \def\byear#1{#1}\fi
\ifx\bissue    \undefined \def\bissue#1{#1}\fi
\ifx\bfpage    \undefined \def\bfpage#1{#1}\fi
\ifx\blpage    \undefined \def\blpage #1{#1}\fi
\ifx\burl      \undefined \def\burl#1{\textsf{#1}}\fi
\ifx\href      \undefined \def\href#1#2{\textsf{#2}}\fi
\ifx\betal     \undefined \def\betal{\textit{et al.}}\fi
\ifx\bctitle   \undefined \def\bctitle#1{#1}\fi
\ifx\beditor   \undefined \def\beditor#1{#1}\fi
\ifx\bbtitle   \undefined \def\bbtitle#1{\textit{#1}}\fi
\ifx\bedition  \undefined \def\bedition#1{#1}\fi
\ifx\bseriesno \undefined \def\bseriesno#1{\textbf{#1}}\fi
\ifx\blocation \undefined \def\blocation#1{#1}\fi
\ifx\bsertitle \undefined \def\bsertitle#1{\textit{#1}}\fi
\ifx\bsnm      \undefined \def\bsnm#1{#1}\fi
\ifx\bsuffix   \undefined \def\bsuffix#1{#1}\fi
\ifx\bparticle \undefined \def\bparticle#1{#1}\fi
\ifx\barticle  \undefined \def\barticle#1{}\fi
\ifx\binstitute  \undefined \def\binstitute#1{#1}\fi
\ifx\bpublisher  \undefined \def\bpublisher#1{#1}\fi
\ifx\doiurl    \undefined
  \def\doiurl#1{\href{http://dx.doi.org/#1}{\textsf{DOI}}}\fi
\ifx\arxivurl  \undefined
  \def\arxivurl#1{\href{http://arxiv.org/abs/#1}{\textsf{arXiv}}}\fi
\ifx\adsurl    \undefined
  \def\adsurl#1{\href{http://adsabs.harvard.edu/abs/#1}{\textsf{ADS}}}\fi
\ifx\botherref \undefined \def\botherref#1{}\fi
\ifx\url       \undefined \def\url#1{\textsf{#1}}\fi
\ifx\bchapter  \undefined \def\bchapter#1{}\fi
\ifx\bbook     \undefined \def\bbook#1{}\fi
\ifx\bcomment  \undefined \def\bcomment#1{#1}\fi
\ifx\oauthor   \undefined \def\oauthor#1{#1}\fi
\ifx\citeauthoryear \undefined\def \citeauthoryear#1{#1}\fi
\ifx\endbibitem\undefined \def\endbibitem{}\fi
\ifx\bconflocation  \undefined \def\bconflocation#1{#1} \fi

\bibitem[\protect\citeauthoryear{Aschwanden}{2011}]{aschwanden_self-organized_2011}
\begin{bbook}
\bauthor{\bsnm{Aschwanden}, \binits{M.}}:
\byear{2011},
\bbtitle{Self-{Organized} {Criticality} in {Astrophysics}: {The} {Statistics}
  of {Nonlinear} {Processes} in the {Universe}},
\bsertitle{Astronomy and {Planetary} {Sciences}},
\bpublisher{Springer},
\blocation{Berlin Heidelberg}.
\bisbn{978-3-642-15000-5}.
\doiurl{10.1007/978-3-642-15001-2}.
\burl{https://www.springer.com/gp/book/9783642150005}.
\end{bbook}
\endbibitem

\bibitem[\protect\citeauthoryear{Aschwanden}{2013}]{aschwanden_self-organized_2013}
\begin{botherref}
\oauthor{\bsnm{Aschwanden}, \binits{M.J.}}:
2013,
Self-{Organized} {Criticality} {Systems}.
\textit{Self-Organized Criticality Systems, Edited by M.J. Aschwanden. e-book
  published by Open Academic Press, Berlin, Warsaw, 2013, 483pp.}
\url{http://adsabs.harvard.edu/abs/2013socs.book.....A}.
\end{botherref}
\endbibitem

\bibitem[\protect\citeauthoryear{Aschwanden and
  Freeland}{2012}]{aschwanden_automated_2012}
\begin{barticle}
\bauthor{\bsnm{Aschwanden}, \binits{M.J.}},
\bauthor{\bsnm{Freeland}, \binits{S.L.}}:
\byear{2012},
\batitle{Automated {Solar} {Flare} {Statistics} in {Soft} {X}-{Rays} over 37
  {Years} of {GOES} {Observations}: {The} {Invariance} of {Self}-organized
  {Criticality} during {Three} {Solar} {Cycles}}.
\bjtitle{The Astrophysical Journal}
\bvolume{754},
\bfpage{112}.
\bcomment{ADS Bibcode: 2012ApJ...754..112A}.
\doiurl{10.1088/0004-637X/754/2/112}.
\burl{https://ui.adsabs.harvard.edu/abs/2012ApJ...754..112A}.
\end{barticle}
\endbibitem

\bibitem[\protect\citeauthoryear{Aschwanden, Stern, and
  Güdel}{2008}]{aschwanden_scaling_2008}
\begin{barticle}
\bauthor{\bsnm{Aschwanden}, \binits{M.J.}},
\bauthor{\bsnm{Stern}, \binits{R.A.}},
\bauthor{\bsnm{Güdel}, \binits{M.}}:
\byear{2008},
\batitle{Scaling laws of solar and stellar flares}.
\bjtitle{The Astrophysical Journal}
\bvolume{672}(\bissue{1}),
\bfpage{659}.
\bcomment{arXiv: 0710.2563}.
\doiurl{10.1086/523926}.
\burl{http://arxiv.org/abs/0710.2563}.
\end{barticle}
\endbibitem

\bibitem[\protect\citeauthoryear{Aschwanden
  \textit{et~al.}}{2016}]{aschwanden_25_2016}
\begin{barticle}
\bauthor{\bsnm{Aschwanden}, \binits{M.J.}},
\bauthor{\bsnm{Crosby}, \binits{N.B.}},
\bauthor{\bsnm{Dimitropoulou}, \binits{M.}},
\bauthor{\bsnm{Georgoulis}, \binits{M.K.}},
\bauthor{\bsnm{Hergarten}, \binits{S.}},
\bauthor{\bsnm{McAteer}, \binits{J.}},
\bauthor{\bsnm{Milovanov}, \binits{A.V.}},
\bauthor{\bsnm{Mineshige}, \binits{S.}},
\bauthor{\bsnm{Morales}, \binits{L.}},
\bauthor{\bsnm{Nishizuka}, \binits{N.}},
\bauthor{\bsnm{Pruessner}, \binits{G.}},
\bauthor{\bsnm{Sanchez}, \binits{R.}},
\bauthor{\bsnm{Sharma}, \binits{A.S.}},
\bauthor{\bsnm{Strugarek}, \binits{A.}},
\bauthor{\bsnm{Uritsky}, \binits{V.}}:
\byear{2016},
\batitle{25 {Years} of {Self}-{Organized} {Criticality}: {Solar} and
  {Astrophysics}}.
\bjtitle{Space Science Reviews}
\bvolume{198}(\bissue{1-4}),
\bfpage{47}.
\doiurl{10.1007/s11214-014-0054-6}.
\burl{http://link.springer.com/10.1007/s11214-014-0054-6}.
\end{barticle}
\endbibitem

\bibitem[\protect\citeauthoryear{Bak, Tang, and
  Wiesenfeld}{1987}]{bak_self-organized_1987}
\begin{barticle}
\bauthor{\bsnm{Bak}, \binits{P.}},
\bauthor{\bsnm{Tang}, \binits{C.}},
\bauthor{\bsnm{Wiesenfeld}, \binits{K.}}:
\byear{1987},
\batitle{Self-organized criticality: {An} explanation of the 1/f noise}.
\bjtitle{Physical Review Letters}
\bvolume{59}(\bissue{4}),
\bfpage{381}.
\bcomment{Publisher: American Physical Society}.
\doiurl{10.1103/PhysRevLett.59.381}.
\burl{https://link.aps.org/doi/10.1103/PhysRevLett.59.381}.
\end{barticle}
\endbibitem

\bibitem[\protect\citeauthoryear{{Barnes} and
  {Leka}}{2008}]{2008ApJ...688L.107B}
\begin{barticle}
\bauthor{\bsnm{{Barnes}}, \binits{G.}},
\bauthor{\bsnm{{Leka}}, \binits{K.D.}}:
\byear{2008},
\batitle{{Evaluating the Performance of Solar Flare Forecasting Methods}}.
\bjtitle{\apjl}
\bvolume{688}(\bissue{2}),
\bfpage{L107}.
\doiurl{10.1086/595550}.
\adsurl{https://ui.adsabs.harvard.edu/abs/2008ApJ...688L.107B}.
\end{barticle}
\endbibitem

\bibitem[\protect\citeauthoryear{Barnes
  \textit{et~al.}}{2016}]{barnes_comparison_2016}
\begin{barticle}
\bauthor{\bsnm{Barnes}, \binits{G.}},
\bauthor{\bsnm{Leka}, \binits{K.D.}},
\bauthor{\bsnm{Schrijver}, \binits{C.J.}},
\bauthor{\bsnm{Colak}, \binits{T.}},
\bauthor{\bsnm{Qahwaji}, \binits{R.}},
\bauthor{\bsnm{Ashamari}, \binits{O.W.}},
\bauthor{\bsnm{Yuan}, \binits{Y.}},
\bauthor{\bsnm{Zhang}, \binits{J.}},
\bauthor{\bsnm{McAteer}, \binits{R.T.J.}},
\bauthor{\bsnm{Bloomfield}, \binits{D.S.}},
\bauthor{\bsnm{Higgins}, \binits{P.A.}},
\bauthor{\bsnm{Gallagher}, \binits{P.T.}},
\bauthor{\bsnm{Falconer}, \binits{D.A.}},
\bauthor{\bsnm{Georgoulis}, \binits{M.K.}},
\bauthor{\bsnm{Wheatland}, \binits{M.S.}},
\bauthor{\bsnm{Balch}, \binits{C.}},
\bauthor{\bsnm{Dunn}, \binits{T.}},
\bauthor{\bsnm{Wagner}, \binits{E.L.}}:
\byear{2016},
\batitle{A {Comparison} of {Flare} {Forecasting} {Methods}, {I}: {Results} from
  the "{All}-{Clear}" {Workshop}}.
\bjtitle{The Astrophysical Journal}
\bvolume{829}(\bissue{2}),
\bfpage{89}.
\bcomment{arXiv: 1608.06319}.
\doiurl{10.3847/0004-637X/829/2/89}.
\burl{http://arxiv.org/abs/1608.06319}.
\end{barticle}
\endbibitem

\bibitem[\protect\citeauthoryear{Bélanger, Vincent, and
  Charbonneau}{2007}]{belanger_predicting_2007}
\begin{barticle}
\bauthor{\bsnm{Bélanger}, \binits{E.}},
\bauthor{\bsnm{Vincent}, \binits{A.}},
\bauthor{\bsnm{Charbonneau}, \binits{P.}}:
\byear{2007},
\batitle{Predicting {Solar} {Flares} by {Data} {Assimilation} in {Avalanche}
  {Models}}.
\bjtitle{Solar Physics}
\bvolume{245}(\bissue{1}),
\bfpage{141}.
\doiurl{10.1007/s11207-007-9009-3}.
\burl{https://doi.org/10.1007/s11207-007-9009-3}.
\end{barticle}
\endbibitem

\bibitem[\protect\citeauthoryear{Charbonneau
  \textit{et~al.}}{2001}]{charbonneau_avalanche_2001}
\begin{botherref}
\oauthor{\bsnm{Charbonneau}, \binits{P.}},
\oauthor{\bsnm{Mcintosh}, \binits{S.}},
\oauthor{\bsnm{Liu}, \binits{H.}},
\oauthor{\bsnm{Bogdan}, \binits{J.}}:
2001,
Avalanche models for solar flares ({Invited} {Review}).
\textit{Solar Physics},
321.
\doiurl{10.1023/A:1013301521745}.
\url{https://opensky.ucar.edu/islandora/object/articles\%3A8921/}.
\end{botherref}
\endbibitem

\bibitem[\protect\citeauthoryear{Cheung
  \textit{et~al.}}{2019}]{cheung_comprehensive_2019}
\begin{barticle}
\bauthor{\bsnm{Cheung}, \binits{M.C.M.}},
\bauthor{\bsnm{Rempel}, \binits{M.}},
\bauthor{\bsnm{Chintzoglou}, \binits{G.}},
\bauthor{\bsnm{Chen}, \binits{F.}},
\bauthor{\bsnm{Testa}, \binits{P.}},
\bauthor{\bsnm{Martínez-Sykora}, \binits{J.}},
\bauthor{\bsnm{Sainz~Dalda}, \binits{A.}},
\bauthor{\bsnm{DeRosa}, \binits{M.L.}},
\bauthor{\bsnm{Malanushenko}, \binits{A.}},
\bauthor{\bsnm{Hansteen}, \binits{V.}},
\bauthor{\bsnm{De~Pontieu}, \binits{B.}},
\bauthor{\bsnm{Carlsson}, \binits{M.}},
\bauthor{\bsnm{Gudiksen}, \binits{B.}},
\bauthor{\bsnm{McIntosh}, \binits{S.W.}}:
\byear{2019},
\batitle{A comprehensive three-dimensional radiative magnetohydrodynamic
  simulation of a solar flare}.
\bjtitle{Nature Astronomy}
\bvolume{3},
\bfpage{160}.
\doiurl{10.1038/s41550-018-0629-3}.
\burl{http://adsabs.harvard.edu/abs/2019NatAs...3..160C}.
\end{barticle}
\endbibitem

\bibitem[\protect\citeauthoryear{Dennis}{1985}]{dennis_solar_1985}
\begin{barticle}
\bauthor{\bsnm{Dennis}, \binits{B.R.}}:
\byear{1985},
\batitle{Solar hard {X}-ray bursts}.
\bjtitle{Solar Physics}
\bvolume{100},
\bfpage{465}.
\doiurl{10.1007/BF00158441}.
\burl{http://adsabs.harvard.edu/abs/1985SoPh..100..465D}.
\end{barticle}
\endbibitem

\bibitem[\protect\citeauthoryear{Farhang, Safari, and
  Wheatland}{2018}]{farhang_principle_2018}
\begin{barticle}
\bauthor{\bsnm{Farhang}, \binits{N.}},
\bauthor{\bsnm{Safari}, \binits{H.}},
\bauthor{\bsnm{Wheatland}, \binits{M.S.}}:
\byear{2018},
\batitle{Principle of {Minimum} {Energy} in {Magnetic} {Reconnection} in a
  {Self}-organized {Critical} {Model} for {Solar} {Flares}}.
\bjtitle{The Astrophysical Journal}
\bvolume{859}(\bissue{1}),
\bfpage{41}.
\bcomment{Publisher: American Astronomical Society}.
\doiurl{10.3847/1538-4357/aac01b}.
\burl{https://doi.org/10.3847\%2F1538-4357\%2Faac01b}.
\end{barticle}
\endbibitem

\bibitem[\protect\citeauthoryear{Farhang, Wheatland, and
  Safari}{2019}]{farhang_energy_2019}
\begin{barticle}
\bauthor{\bsnm{Farhang}, \binits{N.}},
\bauthor{\bsnm{Wheatland}, \binits{M.S.}},
\bauthor{\bsnm{Safari}, \binits{H.}}:
\byear{2019},
\batitle{Energy {Balance} in {Avalanche} {Models} for {Solar} {Flares}}.
\bjtitle{The Astrophysical Journal}
\bvolume{883}(\bissue{1}),
\bfpage{L20}.
\bcomment{Publisher: American Astronomical Society}.
\doiurl{10.3847/2041-8213/ab40c3}.
\burl{https://doi.org/10.3847\%2F2041-8213\%2Fab40c3}.
\end{barticle}
\endbibitem

\bibitem[\protect\citeauthoryear{Hung
  \textit{et~al.}}{2017}]{hung_variational_2017}
\begin{barticle}
\bauthor{\bsnm{Hung}, \binits{C.P.}},
\bauthor{\bsnm{Brun}, \binits{A.S.}},
\bauthor{\bsnm{Fournier}, \binits{A.}},
\bauthor{\bsnm{Jouve}, \binits{L.}},
\bauthor{\bsnm{Talagrand}, \binits{O.}},
\bauthor{\bsnm{Zakari}, \binits{M.}}:
\byear{2017},
\batitle{Variational estimation of the large scale time dependent meridional
  circulation in the {Sun}: proofs of concept with a solar mean field dynamo
  model}.
\bjtitle{The Astrophysical Journal}
\bvolume{849}(\bissue{2}),
\bfpage{160}.
\bcomment{arXiv: 1710.02114}.
\doiurl{10.3847/1538-4357/aa91d1}.
\burl{http://arxiv.org/abs/1710.02114}.
\end{barticle}
\endbibitem

\bibitem[\protect\citeauthoryear{Jensen}{1998}]{jensen_self-organized_1998}
\begin{bbook}
\bauthor{\bsnm{Jensen}, \binits{H.J.}}:
\byear{1998},
\bbtitle{Self-{Organized} {Criticality}: {Emergent} {Complex} {Behavior} in
  {Physical} and {Biological} {Systems}},
\bedition{1 edition} edn.
\bpublisher{Cambridge University Press},
\blocation{Cambridge ; New York}.
\bisbn{978-0-521-48371-1}.
\end{bbook}
\endbibitem

\bibitem[\protect\citeauthoryear{Kalnay}{2002}]{kalnay_atmospheric_2002}
\begin{bbook}
\bauthor{\bsnm{Kalnay}, \binits{E.}}:
\byear{2002},
\bbtitle{Atmospheric {Modeling}, {Data} {Assimilation} and {Predictability}},
\bedition{1 edition} edn.
\bpublisher{Cambridge University Press},
\blocation{New York}.
\bisbn{978-0-521-79629-3}.
\end{bbook}
\endbibitem

\bibitem[\protect\citeauthoryear{Leka
  \textit{et~al.}}{2019a}]{leka_comparison_2019}
\begin{barticle}
\bauthor{\bsnm{Leka}, \binits{K.D.}},
\bauthor{\bsnm{Park}, \binits{S.-H.}},
\bauthor{\bsnm{Kusano}, \binits{K.}},
\bauthor{\bsnm{Andries}, \binits{J.}},
\bauthor{\bsnm{Barnes}, \binits{G.}},
\bauthor{\bsnm{Bingham}, \binits{S.}},
\bauthor{\bsnm{Bloomfield}, \binits{D.S.}},
\bauthor{\bsnm{McCloskey}, \binits{A.E.}},
\bauthor{\bsnm{Delouille}, \binits{V.}},
\bauthor{\bsnm{Falconer}, \binits{D.}},
\bauthor{\bsnm{Gallagher}, \binits{P.T.}},
\bauthor{\bsnm{Georgoulis}, \binits{M.K.}},
\bauthor{\bsnm{Kubo}, \binits{Y.}},
\bauthor{\bsnm{Lee}, \binits{K.}},
\bauthor{\bsnm{Lee}, \binits{S.}},
\bauthor{\bsnm{Lobzin}, \binits{V.}},
\bauthor{\bsnm{Mun}, \binits{J.}},
\bauthor{\bsnm{Murray}, \binits{S.A.}},
\bauthor{\bsnm{Nageem}, \binits{T.A.M.H.}},
\bauthor{\bsnm{Qahwaji}, \binits{R.}},
\bauthor{\bsnm{Sharpe}, \binits{M.}},
\bauthor{\bsnm{Steenburgh}, \binits{R.A.}},
\bauthor{\bsnm{Steward}, \binits{G.}},
\bauthor{\bsnm{Terkildsen}, \binits{M.}}:
\byear{2019}a,
\batitle{A {Comparison} of {Flare} {Forecasting} {Methods}. {II}. {Benchmarks},
  {Metrics}, and {Performance} {Results} for {Operational} {Solar} {Flare}
  {Forecasting} {Systems}}.
\bjtitle{The Astrophysical Journal Supplement Series}
\bvolume{243}(\bissue{2}),
\bfpage{36}.
\bcomment{Publisher: American Astronomical Society}.
\doiurl{10.3847/1538-4365/ab2e12}.
\burl{https://doi.org/10.3847\%2F1538-4365\%2Fab2e12}.
\end{barticle}
\endbibitem

\bibitem[\protect\citeauthoryear{Leka
  \textit{et~al.}}{2019b}]{leka_comparison_2019-1}
\begin{barticle}
\bauthor{\bsnm{Leka}, \binits{K.D.}},
\bauthor{\bsnm{Park}, \binits{S.-H.}},
\bauthor{\bsnm{Kusano}, \binits{K.}},
\bauthor{\bsnm{Andries}, \binits{J.}},
\bauthor{\bsnm{Barnes}, \binits{G.}},
\bauthor{\bsnm{Bingham}, \binits{S.}},
\bauthor{\bsnm{Bloomfield}, \binits{D.S.}},
\bauthor{\bsnm{McCloskey}, \binits{A.E.}},
\bauthor{\bsnm{Delouille}, \binits{V.}},
\bauthor{\bsnm{Falconer}, \binits{D.}},
\bauthor{\bsnm{Gallagher}, \binits{P.T.}},
\bauthor{\bsnm{Georgoulis}, \binits{M.K.}},
\bauthor{\bsnm{Kubo}, \binits{Y.}},
\bauthor{\bsnm{Lee}, \binits{K.}},
\bauthor{\bsnm{Lee}, \binits{S.}},
\bauthor{\bsnm{Lobzin}, \binits{V.}},
\bauthor{\bsnm{Mun}, \binits{J.}},
\bauthor{\bsnm{Murray}, \binits{S.A.}},
\bauthor{\bsnm{Nageem}, \binits{T.A.M.H.}},
\bauthor{\bsnm{Qahwaji}, \binits{R.}},
\bauthor{\bsnm{Sharpe}, \binits{M.}},
\bauthor{\bsnm{Steenburgh}, \binits{R.}},
\bauthor{\bsnm{Steward}, \binits{G.}},
\bauthor{\bsnm{Terkildsen}, \binits{M.}}:
\byear{2019}b,
\batitle{A {Comparison} of {Flare} {Forecasting} {Methods}. {III}. {Systematic}
  {Behaviors} of {Operational} {Solar} {Flare} {Forecasting} {Systems}}.
\bjtitle{The Astrophysical Journal}
\bvolume{881}(\bissue{2}),
\bfpage{101}.
\bcomment{arXiv: 1907.02909}.
\doiurl{10.3847/1538-4357/ab2e11}.
\burl{http://arxiv.org/abs/1907.02909}.
\end{barticle}
\endbibitem

\bibitem[\protect\citeauthoryear{Liu \textit{et~al.}}{2006}]{liu_energy_2006}
\begin{barticle}
\bauthor{\bsnm{Liu}, \binits{W.W.}},
\bauthor{\bsnm{Charbonneau}, \binits{P.}},
\bauthor{\bsnm{Thibault}, \binits{K.}},
\bauthor{\bsnm{Morales}, \binits{L.}}:
\byear{2006},
\batitle{Energy avalanches in the central plasma sheet}.
\bjtitle{Geophysical Research Letters}
\bvolume{33}(\bissue{19}).
\bcomment{\_eprint:
  https://agupubs.onlinelibrary.wiley.com/doi/pdf/10.1029/2006GL027282}.
\doiurl{https://doi.org/10.1029/2006GL027282}.
\burl{https://agupubs.onlinelibrary.wiley.com/doi/abs/10.1029/2006GL027282}.
\end{barticle}
\endbibitem

\bibitem[\protect\citeauthoryear{Lu and Hamilton}{1991}]{lu_avalanches_1991}
\begin{barticle}
\bauthor{\bsnm{Lu}, \binits{E.T.}},
\bauthor{\bsnm{Hamilton}, \binits{R.J.}}:
\byear{1991},
\batitle{Avalanches and the distribution of solar flares}.
\bjtitle{The Astrophysical Journal Letters}
\bvolume{380},
\bfpage{L89}.
\doiurl{10.1086/186180}.
\burl{http://adsabs.harvard.edu/abs/1991ApJ...380L..89L}.
\end{barticle}
\endbibitem

\bibitem[\protect\citeauthoryear{Lu \textit{et~al.}}{1993}]{lu_solar_1993}
\begin{barticle}
\bauthor{\bsnm{Lu}, \binits{E.T.}},
\bauthor{\bsnm{Hamilton}, \binits{R.J.}},
\bauthor{\bsnm{McTiernan}, \binits{J.M.}},
\bauthor{\bsnm{Bromund}, \binits{K.R.}}:
\byear{1993},
\batitle{Solar flares and avalanches in driven dissipative systems}.
\bjtitle{The Astrophysical Journal}
\bvolume{412},
\bfpage{841}.
\doiurl{10.1086/172966}.
\burl{http://adsabs.harvard.edu/abs/1993ApJ...412..841L}.
\end{barticle}
\endbibitem

\bibitem[\protect\citeauthoryear{Morales and
  Charbonneau}{2008}]{morales_self-organized_2008}
\begin{barticle}
\bauthor{\bsnm{Morales}, \binits{L.}},
\bauthor{\bsnm{Charbonneau}, \binits{P.}}:
\byear{2008},
\batitle{Self-organized {Critical} {Model} of {Energy} {Release} in an
  {Idealized} {Coronal} {Loop}}.
\bjtitle{The Astrophysical Journal}
\bvolume{682}(\bissue{1}),
\bfpage{654}.
\bcomment{Publisher: IOP Publishing}.
\doiurl{10.1086/588274}.
\burl{https://iopscience.iop.org/article/10.1086/588274/meta}.
\end{barticle}
\endbibitem

\bibitem[\protect\citeauthoryear{{Morales} and
  {Santos}}{2020}]{2020SoPh..295..155M}
\begin{barticle}
\bauthor{\bsnm{{Morales}}, \binits{L.F.}},
\bauthor{\bsnm{{Santos}}, \binits{N.A.}}:
\byear{2020},
\batitle{{Predicting Extreme Solar Flare Events Using Lu and Hamilton Avalanche
  Model}}.
\bjtitle{\solphys}
\bvolume{295}(\bissue{11}),
\bfpage{155}.
\doiurl{10.1007/s11207-020-01713-0}.
\adsurl{https://ui.adsabs.harvard.edu/abs/2020SoPh..295..155M}.
\end{barticle}
\endbibitem

\bibitem[\protect\citeauthoryear{Namekata
  \textit{et~al.}}{2017}]{namekata_statistical_2017}
\begin{barticle}
\bauthor{\bsnm{Namekata}, \binits{K.}},
\bauthor{\bsnm{Sakaue}, \binits{T.}},
\bauthor{\bsnm{Watanabe}, \binits{K.}},
\bauthor{\bsnm{Asai}, \binits{A.}},
\bauthor{\bsnm{Maehara}, \binits{H.}},
\bauthor{\bsnm{Notsu}, \binits{Y.}},
\bauthor{\bsnm{Notsu}, \binits{S.}},
\bauthor{\bsnm{Honda}, \binits{S.}},
\bauthor{\bsnm{Ishii}, \binits{T.T.}},
\bauthor{\bsnm{Ikuta}, \binits{K.}},
\bauthor{\bsnm{Nogami}, \binits{D.}},
\bauthor{\bsnm{Shibata}, \binits{K.}}:
\byear{2017},
\batitle{Statistical {Studies} of {Solar} {White}-light {Flares} and
  {Comparisons} with {Superflares} on {Solar}-type {Stars}}.
\bjtitle{The Astrophysical Journal}
\bvolume{851},
\bfpage{91}.
\doiurl{10.3847/1538-4357/aa9b34}.
\burl{http://adsabs.harvard.edu/abs/2017ApJ...851...91N}.
\end{barticle}
\endbibitem

\bibitem[\protect\citeauthoryear{Norman
  \textit{et~al.}}{2001}]{norman_waiting-time_2001}
\begin{barticle}
\bauthor{\bsnm{Norman}, \binits{J.P.}},
\bauthor{\bsnm{Charbonneau}, \binits{P.}},
\bauthor{\bsnm{McIntosh}, \binits{S.W.}},
\bauthor{\bsnm{Liu}, \binits{H.-L.}}:
\byear{2001},
\batitle{Waiting-{Time} {Distributions} in {Lattice} {Models} of {Solar}
  {Flares}}.
\bjtitle{The Astrophysical Journal}
\bvolume{557},
\bfpage{891}.
\doiurl{10.1086/321678}.
\burl{http://adsabs.harvard.edu/abs/2001ApJ...557..891N}.
\end{barticle}
\endbibitem

\bibitem[\protect\citeauthoryear{Olami, Feder, and
  Christensen}{1992}]{olami_self-organized_1992}
\begin{barticle}
\bauthor{\bsnm{Olami}, \binits{Z.}},
\bauthor{\bsnm{Feder}, \binits{H.J.S.}},
\bauthor{\bsnm{Christensen}, \binits{K.}}:
\byear{1992},
\batitle{Self-organized criticality in a continuous, nonconservative cellular
  automaton modeling earthquakes}.
\bjtitle{Physical Review Letters}
\bvolume{68}(\bissue{8}),
\bfpage{1244}.
\bcomment{Publisher: American Physical Society}.
\doiurl{10.1103/PhysRevLett.68.1244}.
\burl{https://link.aps.org/doi/10.1103/PhysRevLett.68.1244}.
\end{barticle}
\endbibitem

\bibitem[\protect\citeauthoryear{Parker}{1988}]{parker_nanoflares_1988}
\begin{barticle}
\bauthor{\bsnm{Parker}, \binits{E.N.}}:
\byear{1988},
\batitle{Nanoflares and the solar {X}-ray corona}.
\bjtitle{The Astrophysical Journal}
\bvolume{330},
\bfpage{474}.
\doiurl{10.1086/166485}.
\burl{http://adsabs.harvard.edu/abs/1988ApJ...330..474P}.
\end{barticle}
\endbibitem

\bibitem[\protect\citeauthoryear{Pascual and
  Hascoëet}{2006}]{pascual_extension_2006}
\begin{bchapter}
\bauthor{\bsnm{Pascual}, \binits{V.}},
\bauthor{\bsnm{Hascoëet}, \binits{L.}}:
\byear{2006},
\bctitle{Extension of {TAPENADE} toward {Fortran} 95}.
In: \beditor{\bsnm{Bücker}, \binits{M.}},
\beditor{\bsnm{Corliss}, \binits{G.}},
\beditor{\bsnm{Naumann}, \binits{U.}},
\beditor{\bsnm{Hovland}, \binits{P.}},
\beditor{\bsnm{Norris}, \binits{B.}} (eds.)
\bbtitle{Automatic {Differentiation}: {Applications}, {Theory}, and
  {Implementations}},
\bsertitle{Lecture {Notes} in {Computational} {Science} and {Engineering}},
\bpublisher{Springer},
\blocation{Berlin, Heidelberg},
\bfpage{171}.
\bisbn{978-3-540-28438-3}.
\doiurl{10.1007/3-540-28438-9_15}.
\end{bchapter}
\endbibitem

\bibitem[\protect\citeauthoryear{Press
  \textit{et~al.}}{1992}]{press_numerical_1992}
\begin{bbook}
\bauthor{\bsnm{Press}, \binits{W.H.}},
\bauthor{\bsnm{Flannery}, \binits{B.P.}},
\bauthor{\bsnm{Teukolsky}, \binits{S.A.}},
\bauthor{\bsnm{Vetterling}, \binits{W.T.}}:
\byear{1992},
\bbtitle{Numerical {Recipes} in {FORTRAN} 77: {Volume} 1, {Volume} 1 of
  {Fortran} {Numerical} {Recipes}: {The} {Art} of {Scientific} {Computing}}.
\end{bbook}
\endbibitem

\bibitem[\protect\citeauthoryear{Strugarek and
  Charbonneau}{2014}]{strugarek_predictive_2014}
\begin{barticle}
\bauthor{\bsnm{Strugarek}, \binits{A.}},
\bauthor{\bsnm{Charbonneau}, \binits{P.}}:
\byear{2014},
\batitle{Predictive {Capabilities} of {Avalanche} {Models} for {Solar}
  {Flares}}.
\bjtitle{Solar Physics}
\bvolume{289}(\bissue{11}),
\bfpage{4137}.
\doiurl{10.1007/s11207-014-0570-2}.
\burl{https://doi.org/10.1007/s11207-014-0570-2}.
\end{barticle}
\endbibitem

\bibitem[\protect\citeauthoryear{Strugarek
  \textit{et~al.}}{2014}]{strugarek_deterministically_2014}
\begin{barticle}
\bauthor{\bsnm{Strugarek}, \binits{A.}},
\bauthor{\bsnm{Charbonneau}, \binits{P.}},
\bauthor{\bsnm{Joseph}, \binits{R.}},
\bauthor{\bsnm{Pirot}, \binits{D.}}:
\byear{2014},
\batitle{Deterministically {Driven} {Avalanche} {Models} of {Solar} {Flares}}.
\bjtitle{Solar Physics}
\bvolume{289}(\bissue{8}),
\bfpage{2993}.
\doiurl{10.1007/s11207-014-0509-7}.
\burl{https://doi.org/10.1007/s11207-014-0509-7}.
\end{barticle}
\endbibitem

\bibitem[\protect\citeauthoryear{Strugarek
  \textit{et~al.}}{2017}]{strugarek_sandpile_2017}
\begin{barticle}
\bauthor{\bsnm{Strugarek}, \binits{A.}},
\bauthor{\bsnm{Brun}, \binits{A.S.}},
\bauthor{\bsnm{Charbonneau}, \binits{P.}},
\bauthor{\bsnm{Vilmer}, \binits{N.}}:
\byear{2017},
\batitle{Sandpile {Models} and {Solar} {Flares}: {Eigenfunction}
  {Decomposition} for {Data} {Assimilation}}.
\bjtitle{Proceedings of the International Astronomical Union}
\bvolume{13}(\bissue{S335}),
\bfpage{250}.
\bcomment{Publisher: Cambridge University Press}.
\doiurl{10.1017/S1743921317007244}.
\burl{https://www.cambridge.org/core/journals/proceedings-of-the-international-astronomical-union/article/sandpile-models-and-solar-flares-eigenfunction-decomposition-for-data-assimilation/6217C836E893573DEE272CB39A5CCEE5}.
\end{barticle}
\endbibitem

\bibitem[\protect\citeauthoryear{Vallières‐Nollet
  \textit{et~al.}}{2010}]{vallieresnollet_dual_2010}
\begin{barticle}
\bauthor{\bsnm{Vallières‐Nollet}, \binits{M.-A.}},
\bauthor{\bsnm{Charbonneau}, \binits{P.}},
\bauthor{\bsnm{Uritsky}, \binits{V.}},
\bauthor{\bsnm{Donovan}, \binits{E.}},
\bauthor{\bsnm{Liu}, \binits{W.}}:
\byear{2010},
\batitle{Dual scaling for self-organized critical models of the magnetosphere}.
\bjtitle{Journal of Geophysical Research: Space Physics}
\bvolume{115}(\bissue{A12}).
\bcomment{\_eprint:
  https://agupubs.onlinelibrary.wiley.com/doi/pdf/10.1029/2010JA015641}.
\doiurl{https://doi.org/10.1029/2010JA015641}.
\burl{https://agupubs.onlinelibrary.wiley.com/doi/abs/10.1029/2010JA015641}.
\end{barticle}
\endbibitem

\bibitem[\protect\citeauthoryear{{Watkins}
  \textit{et~al.}}{2016}]{2016SSRv..198....3W}
\begin{barticle}
\bauthor{\bsnm{{Watkins}}, \binits{N.W.}},
\bauthor{\bsnm{{Pruessner}}, \binits{G.}},
\bauthor{\bsnm{{Chapman}}, \binits{S.C.}},
\bauthor{\bsnm{{Crosby}}, \binits{N.B.}},
\bauthor{\bsnm{{Jensen}}, \binits{H.J.}}:
\byear{2016},
\batitle{{25 Years of Self-organized Criticality: Concepts and Controversies}}.
\bjtitle{\ssr}
\bvolume{198}(\bissue{1-4}),
\bfpage{3}.
\doiurl{10.1007/s11214-015-0155-x}.
\adsurl{https://ui.adsabs.harvard.edu/abs/2016SSRv..198....3W}.
\end{barticle}
\endbibitem

\bibitem[\protect\citeauthoryear{Wheatland}{2000}]{wheatland_origin_2000}
\begin{barticle}
\bauthor{\bsnm{Wheatland}, \binits{M.S.}}:
\byear{2000},
\batitle{The {Origin} of the {Solar} {Flare} {Waiting}-{Time} {Distribution}}.
\bjtitle{The Astrophysical Journal Letters}
\bvolume{536},
\bfpage{L109}.
\doiurl{10.1086/312739}.
\burl{http://adsabs.harvard.edu/abs/2000ApJ...536L.109W}.
\end{barticle}
\endbibitem

\end{thebibliography}

%\begin{thebibliography}{9}

%\bibitem{Charbonneau2001}
%Charbonneau, P., W. McIntosh, S., Liu, H.-L., \& J. Bogdan, T. (2001). Avalanche models for solar flares (Invited Review). 33.

%\bibitem{Lu1991}
%Lu, E. T., \& Hamilton, R. J. (1991). Avalanches and the distribution of solar flares. The astrophysical journal, 380, L89-L92.

%\bibitem{Strugarek2014}
%Strugarek, A., Charbonneau, P., Joseph, R., \& Pirot, D. (2014). Deterministically Driven Avalanche Models of Solar Flares. Solar Physics, 289(8), 2993–3015.

%\bibitem{StrugarekCharbonneau2014}
%Strugarek, A., \& Charbonneau, P. (2014). Predictive Capabilities of Avalanche Models for Solar Flares. Solar Physics, 289(11), 4137–4150. https://doi.org/10.1007/s11207-014-0570-2

%\end{thebibliography}

\end{article} 

\end{document}